\documentclass[preprint2]{aastex}

\usepackage{epsfig}
\newcommand{\be}{\begin{eqnarray}}
\newcommand{\ee}{\end{eqnarray}}
\newcommand{\beq}{\begin{equation}}
\newcommand{\eeq}{\end{equation}}
\def\simless{\mathbin{\lower 3pt\hbox
      {$\rlap{\raise 5pt\hbox{$\char'074$}}\mathchar"7218$}}}
\def\simgreat{\mathbin{\lower 3pt\hbox
      {$\rlap{\raise 5pt\hbox{$\char'076$}}\mathchar"7218$}}} %> or of order

\newcommand{\kvec}{{\mbox{\boldmath $k$}}}
\newcommand{\vevec}{{\mbox{\boldmath $\varepsilon$}}}
\newcommand{\xivec}{{\mbox{\boldmath $\xi$}}}

\newcommand{\Omegavec}{{\mbox{\boldmath $\Omega$}}}
\newcommand{\xvec}{{\mbox{\boldmath $x$}}}
\newcommand{\gvec}{{\mbox{\boldmath $g$}}}

\newcommand{\calFvec}{ {\bf{\cal F}} }
\newcommand{\ve}{\varepsilon}
\newcommand{\grad}{{\mbox{\boldmath $\nabla$}}}
\newcommand{\unit}{{\mbox{\boldmath $e$}}}
\newcommand{\Eunit}{E_{\rm unit}}
\newcommand{\rstar}{ r_{*}}
\newcommand{\calK}{ {\cal K} }

\newcommand{\calI}{ {\cal I} }
\newcommand{\calF}{ {\cal F} }
\newcommand{\calJ}{ {\cal J} }

\newcommand{\zhat}{ \hat{z} }

\begin{document}

% -----------------------------------------------------------
% -----------------------------------------------------------

\title{ Saturation of the R-mode Instability }
\author{Phil Arras$^{1}$, Eanna E. Flanagan$^{2}$, Sharon M. Morsink$^{3}$,\
A. Katrin Schenk$^{4}$, Saul A. Teukolsky$^{2,5}$, Ira Wasserman$^{2}$ }
\affil{$^1$ Canadian Institute for Theoretical Astrophysics, University
of Toronto \\
$^2$ Center for Radiophysics and Space Research, Cornell University \\
$^3$ Theoretical Physics Institute, Department of Physics, 
University of Alberta \\
$^4$ Sloan-Swartz Center for Theoretical Neurobiology, University of California at San Francisco\\
$^5$ Columbia Astrophysics Laboratory }
\authoraddr{$^1$ 60 St. George Street, Ontario M5S 3H8 Canada \\
$^2$ Ithaca, NY 14853\\
$^3$ Edmonton, AB, Canada, T6G 2J1 \\
$^4$ 513 Parnassus, San Francisco, CA 94143-0444 \\
$^5$ 550 W. 120th St., New York, NY 10027 \\
arras@cita.utoronto.ca, flanagan@spacenet.tn.cornell.edu, 
morsink@phys.ualberta.ca,schenk@phy.ucsf.edu,teukolsky@spacenet.tn.cornell.edu,
ira@spacenet.tn.cornel.edu }

% -----------------------------------------------------------
% -----------------------------------------------------------

\begin{abstract}

Rossby waves (r-modes) in rapidly rotating neutron stars are unstable because
of the emission of gravitational radiation. As a result, the stellar rotational
energy is converted into both gravitational waves and r-mode energy.
The saturation level for the r-mode energy is a fundamental parameter 
needed to determine how fast the neutron star spins down, as well as
whether gravitational waves will be detectable.
In this paper, we study saturation
by nonlinear transfer of energy to the sea of stellar ``inertial" oscillation 
modes which arise in rotating stars with negligible buoyancy
and elastic restoring forces.

We present detailed calculations of stellar inertial modes in the WKB
limit, their linear
damping by bulk and shear viscosity, and the nonlinear coupling forces 
among these modes. The saturation amplitude is derived in the extreme
limits of strong or weak driving by radiation reaction, as compared
to the damping rate of low order inertial modes. In the weak driving case,
energy can be stably transferred to a small number of modes, which damp the
energy as heat or neutrinos.
%
% EEF: I modified the following sentence to remove the reference to
% bulk viscosity
%
% Old version:
%
% In the strong driving case, we show that a 
% turbulent cascade develops, with a constant flux of energy to both large
% wavenumber, damped by shear viscosity, and small frequency, damped by 
% bulk viscosity. 
%
In the strong driving case, we show that a turbulent cascade develops,
with a constant flux of energy to large wavenumbers and small
frequencies where it is damped by shear viscosity.

We find the saturation energy
is {\it extremely small}, at least four orders of magnitude smaller than
that found by previous investigators.
We show that the large saturation energy found in the simulations 
of \citet{2001PhRvL..86.1152L,lindblomv2}
is an artifact of their unphysically large
radiation reaction force.
In most physical situations of interest, for either
nascent, rapidly rotating neutron stars, or neutron stars being spun up
by accretion in Low Mass X-ray Binaries (LMXB's), the strong driving limit is
appropriate and the saturation energy is roughly 
$E_{r-mode}/(0.5M\rstar^2 \Omega^2) \simeq  0.1 \gamma_{gr}/\Omega
\simeq 10^{-6} (\nu_{\rm spin}/10^3\,\rm Hz)^5$, where $M$ and $\rstar$ are the
stellar
mass and radius, $\gamma_{gr}$ is the driving rate by gravitational
radiation, $\Omega$ is the angular velocity of the star, and $\nu_{\rm spin}$
is the spin frequency. At such a low saturation amplitude, the characteristic
time for the star to exit the region of r-mode instability is 
$\simgreat 10^{3-4}$ years, depending sensitively on
the instability curve. Although our saturation amplitude is smaller than that
found by previous investigators, it is still sufficiently large to explain
the observed period clustering in LMXB's. 
%
%  EEF: changed the following sentence
%
%  Old version:
% We comment on the prospects for 
% observing gravitational waves either from LMXB's or young neutron stars.
%
We find that the r-mode signal from both newly born neutron stars and
LMXB's in the spin down phase of Levin's limit cycle will be
detectable by enhanced LIGO detectors out to $\sim 
100 - 200$ kpc.

\end{abstract}

\keywords{stars: neutron --- gravitational waves ---
turbulence --- stars: oscillations  }

% -----------------------------------------------------------
% -----------------------------------------------------------

\section{ Introduction } 

What sets the observed spin rates of neutron stars?

Theoretically, we expect neutron stars can rotate
up to $\sim\ 10^3 \,\rm Hz$ without breaking apart
\citep{1994ApJ...423L.117C,1994ApJ...424..823C,
2000ApJ...541.1033F,2000ApJ...528..368H}.
However, for the rapidly accreting, weakly magnetic LMXB's,
oscillations seen during type I X-ray bursts \citep{1996ApJ...469L...9S},
as well as quasi-periodic oscillations \citep{1998mfns.conf..337V},
seem to indicate spin frequencies narrowly clustered near
$300 \,\rm Hz$. If
LMXB's are the progenitors of millisecond pulsars, and the timescale
over which they should be spun up by accretion is only $\sim 10^7{\rm yr}$
for high accretion rates, why aren't more stars
spun up near $10^3 \,\rm Hz$ over their $> 10^9$ year
lifetime? 

%
% EEF: I added Bildsten's paper to the list of references for the
% rmode/LMXB connection in the following paragraph
%

\citet{Wagoner1984} proposed that for weakly magnetic neutron stars,
the spin up torque due to accretion is balanced by spin down
torque from gravitational radiation reaction. There are currently two
distinct models to explain the non-axisymmetric deformation of the star
producing the radiation. The first mechanism involves
mass quadrupole deformations of the neutron star crust 
\citep{1998ApJ...501L..89B,2000MNRAS.319..902U} while the second
involves mass-current quadrupole emission from 
the r-mode instability \citep{1998ApJ...501L..89B,AKS1999,AJKS2000},
which will be examined in detail in this paper.

%
% EEF:  Footnote about gauge freedom inserted in the following paragraph
%

Many young neutron stars associated with supernova remnants also seem to be
spinning slowly, in spite of the theoretical expectation
\citep{2000ApJ...541.1033F,2000ApJ...528..368H} that typical
$8-25\ M_{\odot}$ progenitors lead to neutron stars rotating with 
periods of order \footnote{ This result depends sensitively on the 
poorly understood coupling between the core and envelope of the progenitor.
Angular momentum transport mechanisms due to, for instance, weak  
magnetic fields may decrease the rotation rate of the core prior to collapse.}
$\sim 1\ {\rm msec}$. \citet{kaspi2002} cite the following examples for
the inferred {\it initial} spin period $P_{\rm init}$ and age $T$
of some of the fastest rotators:
the Crab pulsar with $P_{\rm init} = 19\ {\rm msec}$ and $T = 948\ {\rm yr}$;
PSR J0537-6910
in host remnant N157B with $P_{\rm init} \leq 10\ {\rm msec}$ and
$T=5000 \ {\rm yr}$; 
PSR B1951+32 in CTB 80 has $P_{\rm init} \ll 39\ {\rm msec}$ and
$T=10^5 \ {\rm yr}$. The Crab is by far the most certain estimate for 
$P_{\rm init}$, with a known age from the historical supernova and measured
braking index. 
However, \citet{kaspi2002} also note several slow rotators, such as
PSR J1811-1925 in G11.2-0.3 with $P_{\rm init} = 62 \ {\rm msec}$ and age
$T = 2000\ {\rm yr}$. 

The apparent discrepancy between the theoretically expected fast rotation rates
and the observed slow rotation rates could be reconciled if some mechanism
could slow down fast rotators, effectively preventing them from reaching
millisecond spin rates. The r-mode instability is a possible mechanism.
% We now proceed to give background on this effect.

%
% EEF: added a reference to John Friedman's and Sharon's paper in the next
% sentence as John requested.
%
This instability was discovered by \citet{1998ApJ...502..708A}
and \cite{1998ApJ...502..714F}
showed that all rotating, inviscid stars are
unstable because of this general relativistic effect. 
The instability arises when
certain stellar oscillation modes, called Rossby waves (or r-modes),
are driven unstable by the emission of gravitational waves.
As a result, the rotational energy of the star is converted into both mode
energy and gravitational waves, causing the star to spin down.
%
% EEF: added some references to viscosity timescale computations
% in the following sentence as suggested by John Friedman
%
Detailed calculations 
\citep{LOM,AKS1999a,KS1999,1999PhRvD..60f4006L,2000ApJ...529L..33B,
2001MNRAS.324..917L,1999ApJ...521..764L}
show that viscous dissipation by large scale shear, boundary layer shear
at the crust-core interface, and modified URCA bulk viscosity  are
likely insufficient  
to counter this driving in rapidly rotating neutron stars.
However, \citet{lindblomowen2002a} point out an interesting mechanism for
bulk viscosity arising from hyperon interactions which may overcome the
driving. \citet{2001PhRvD..64d4009M} has investigated the effects of magnetic
fields on the boundary layer, finding that large fields can significantly 
increase the damping rate.
Lastly, the work of 
\cite{2001MNRAS.324..917L} shows that damping from the crust-core
boundary layer leads to a double-valued instability curve, which may
explain why LMXB spin frequencies are lower
than those of the millisecond pulsars. 

The instability may be important in two respects. First, r-modes in 
any neutron star rotating faster than some critical rate will
become unstable, causing the star to rapidly spin down. Hence, 
r-modes may set a maximum rotation rate for neutron stars. Second,
the enormous amount of energy radiated in gravitational waves may be
detectable by LIGO.

In section \ref{sec:modecoupling} we review how nonlinear saturation occurs
in the limits of weak and strong driving. We derive formal expressions for the
saturation amplitude, which depend on the microphysical details of the nonlinear
interaction and damping rates.
Section \ref{sec:discrete} contains a review of nonlinear 
coupling of just three oscillation modes, with emphasis on amplitude saturation
by the parametric instability. Section \ref{sec:continuum} reviews amplitude 
saturation by a continuum of modes in which a well defined inertial range
exists. 
In section \ref{sec:whichmodes}, we discuss the modes present  in
rapidly rotating neutron stars, arguing that the
buoyancy and elastic restoring forces are weak compared to the Coriolis force.
We compute WKB inertial eigenmodes in section \ref{sec:inertialmodes}.
The nonlinear coupling coefficients are computed in section \ref{sec:coupling},
and damping rates in section \ref{sec:damping}. The saturation amplitude for
the discrete case is discussed in section \ref{sec:satd}, and the continuum
case in section \ref{sec:satc}. 
Neutron star spin evolution due to the r-mode instability is discussed
in section \ref{sec:evolution}. Our results are compared to those of
previous investigators in section \ref{sec:comparison}. 
%
% EEF: modified following sentence to refer to the new section
%
We discuss the detectability of the gravitational wave signal in section 
\ref{sec:detection}, and give a summary and 
conclusions in section \ref{sec:conclusions}. Two appendices
give detailed calculations of the turbulent cascade for stellar inertial
modes, and the nonlinear force coefficients.

% -----------------------------------------------------------
% -----------------------------------------------------------

\section{ Saturation by Nonlinear Mode Coupling }
\label{sec:modecoupling}

We start by reviewing the equations of motion for the mode amplitudes, 
and then specialize to the weak and strong driving limits.

We will work in a reference frame co-rotating with the star.
Expansion of the fluid displacements, relative to the co-rotating
frame, in terms of the linear
eigenmodes 
\begin{eqnarray}
 \left[ \begin{array}{c}  
        \xivec(t)\\
        {\dot \xivec}(t)
        \end{array} \right] = 
\sum_\alpha q_\alpha(t) \left[ \begin{array}{c}  
        \xivec_\alpha \\
        - i \omega_\alpha \xivec_\alpha
        \end{array} \right]
\end{eqnarray}   
%\begin{eqnarray}
% \left[ \begin{array}{c}
%        \xivec(t)\\
%        {\dot \xivec}(t)
%        \end{array} \right] =
%\sum_\alpha q_{\alpha}(t) \left[ \begin{array}{c}
%        \xivec_\alpha\\
%        - i \omega_\alpha \xivec_\alpha
%        \end{array} \right] +
%q_{\alpha}(t)^* \left[ \begin{array}{c}
%        \xivec_\alpha^*\\
%         i \omega_\alpha \xivec_\alpha^*
%        \end{array} \right].
%\label{goodexpansion1a}
%\end{eqnarray}
%$\xivec(\xvec,t) = \sum_{\alpha} q_{\alpha}(t) \xivec_\alpha(\xvec)$ 
leads to the following system of
coupled oscillator equations for the dimensionless complex amplitudes
$q_\alpha(t)$ \citep{2002PhRvD..65b4001S}:
\be
\dot{q}_\alpha + i \omega_\alpha q_\alpha & = &
\pm \gamma_\alpha q_\alpha 
+ \frac{i}{2} \omega_\alpha \sum_{\beta \gamma}
\kappa_{\alpha \beta \gamma}^* q_\beta^* q_\gamma^*.
\label{eq:amplitudeeqn}
\ee
The left hand side of eq.\ref{eq:amplitudeeqn}
represents an unforced oscillator of
rotating frame frequency $\omega_\alpha$, while the terms on the right hand
side are the driving ($+$) or damping ($-$) term
and the nonlinear term, which is quadratic in $q$.
In our notation, $\kappa_{\alpha \beta \gamma}$
is roughly the ratio of interaction energy to mode energy at
unit amplitude. The rotating frame mode energy is
$E=2\Eunit|q|^2$, where $\Eunit$
is a (arbitrary) unit of energy which we find convenient to set to
$\Eunit=0.5M\rstar^2\Omega^2$
\footnote{ The nonlinear interaction energy
also scales as the rotation energy of the star}.
Here $M$ and $\rstar$ are the stellar mass and radius, and $\Omegavec
=\Omega \unit_z$ is the angular velocity.
%
% EEF: added footnote in the following sentence
%
The sum over modes $\sum_\beta$ involves a sum over the mode 
$(\omega_\beta,\xivec_\beta)$, with amplitude $q_\beta$, as well as its complex
conjugate $(-\omega_\beta,\xivec^*_\beta)$, with amplitude $q^*_\beta$
(see \citealt{2002PhRvD..65b4001S} for a detailed derivation, 
%\footnote{An important aspect of the second order Lagrangian
%perturbation theory which was not discussed by \cite{2002PhRvD..65b4001S} is 
%the gauge freedom discussed by \cite{FS1978}.
%Specifically, for a barotropic star, the Eulerian
%density and velocity perturbations will be preserved to second order
%by transformations of the rotating-frame Lagrangian displacement
%$\xivec$ of
%the form $\xivec({\bf x},t) \to \xivec({\bf
%x},t) + \etavec + (\etavec \cdot \nabla) \xivec + \zetavec$,
%where $\etavec$ and $\zetavec$ satisfy ${\dot \etavec}=0$, $\nabla \cdot (\rho
%\etavec)=0$, ${\dot \zetavec} = 0$, $\nabla \cdot (\rho \zetavec) = \nabla_i
%\nabla_j (\rho \etavec^i \etavec^j)/2$.  We fix this gauge freedom here to
%second order by requiring that the zero-frequency modes discussed
%after Eq. (3.30) of \cite{2002PhRvD..65b4001S} remain unexcited.  Those
%zero-frequency gauge modes are in one-to-one correspondence with the
%inertial modes discussed in this paper.}
but note that
the type of index denoted there by $A$ is denoted here by $\alpha$.)

\subsection{ the discrete limit }
\label{sec:discrete}

\begin{figure*}
\centerline{\psfig{figure=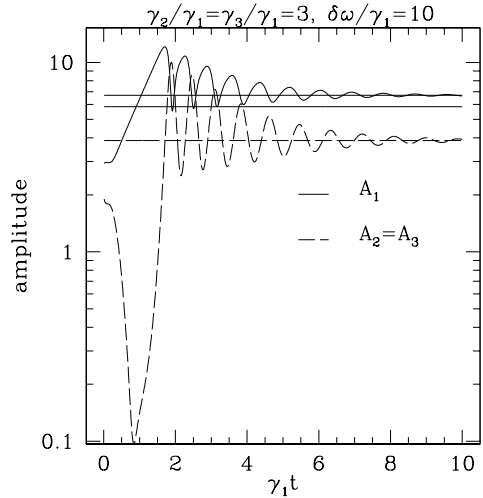,width=0.4\hsize}}
\caption[]{
Saturation in the discrete limit.
Mode amplitudes are evolved in time for the case where
$\gamma_2/\gamma_1=\gamma_3/\gamma_1=3$,
and $\delta \omega / \gamma_1=10$. The solid line is the
driven mode, and the dashed line represents the two daughter modes, which
are identical in this example. The absolute scale is arbitrary.
The amplitudes and phase are started well below their equilibrium values,
which are denoted by the straight upper solid line and the straight
dashed line. The lower straight solid line is the parametric threshold.
Notice that the daughter mode amplitude decreases until the parent exceeds
the threshold, at which point the daughter amplitude rises exponentially.
Qualitatively similar evolutions are obtained for a wide variety of
initial conditions.   The parameters $\gamma_1$, $\gamma_2$,
$\gamma_3$ and $\delta \omega$ were chosen so that the solution would
converge to the equilibrium values.
}
\label{fig:exampleevolution}
\end{figure*}

In the regime where the driving rate of the unstable mode is smaller than
the damping rates of low order modes,
the instability can be saturated by a transferal of energy to
a small number of damped modes.
We will begin by discussing the coupling between the ``parent'' r-mode, and
two damped ``daughter'' modes. Although an idealization, this basic
problem is soluble, and indicates which modes couple most strongly to the 
r-mode. We review the
dynamics of such 3-mode networks, including the parametric
instability, the equilibrium solution, and the linear and nonlinear
stability of the equilibrium solution
\citep{1980PhFl...23.1142W,2001ApJ...546..469W,1985AcA....35....5D,
2000PhRvL..84..622D,abarbanel1993}.

In terms of the real amplitude and
phase variables, defined by $q=A\exp(-i\varphi)$,
the equations for a system of three modes are
\be
\dot{A}_1 & = & \gamma_1 A_1 - \omega_1 \kappa A_{2}A_{3} \sin \varphi
\nonumber \\
\dot{A}_{2} & = & -\gamma_{2} A_{2} - \omega_{2} \kappa A_{3}A_{1} \sin \varphi
\nonumber \\
\dot{A}_{3} & = & -\gamma_{3} A_{3} - \omega_{3} \kappa A_{1}A_{2} \sin \varphi
\nonumber \\
\dot{\varphi} & = & \delta \omega 
%- \kappa \cos\varphi \left[
%\omega_1 \frac{A_{2}A_{3}}{A_1} + \omega_{2} \frac{A_{3}A_1}{A_{2}}
%+ \omega_{3} \frac{A_{1}A_{2}}{A_{3}} 
- \kappa \cos\varphi A_1 A_2 A_3 \left[ \frac{\omega_1}{A_1^2} 
+ \frac{\omega_2}{A_2^2} + \frac{\omega_3}{A_3^2} \right]
\label{eq:3ampeqn}
\ee
where the index 1 refers to the parent and $2$ and $3$
refer to the two daughter modes. We have defined the frequency detuning
$\delta \omega=\omega_1+\omega_{2}+\omega_{3}$,
coupling coefficients $\kappa_{123}=\kappa \exp(-i \delta)$, 
and the relative phase
$\varphi=\delta + \varphi_1+\varphi_{2}+\varphi_{3}$. 
%We have also neglected nonresonant
%quadratic coupling terms such as $\dot{q}_2 \propto q_1^2$ which are 
%unimportant compared to resonant terms. 
The qualitative features
of the time evolution, such as equilibrium and stability,
depend only on the three dimensionless parameters
$(\gamma_{2}+\gamma_{3})/\gamma_1$, $\gamma_{2}/\gamma_{3}$,
and $\delta \omega / \gamma_1$.

The parametric instability \citep{1969mech.book.....L,1985AcA....35....5D,
1996ApJ...466..946K,2001ApJ...546..469W} 
is a mechanism by which the daughter mode
amplitudes will grow exponentially if the parent mode amplitude exceeds a
certain threshold \footnote{ A simple
example of parametric instability is a pendulum in which
the length of the string is being varied periodically.
See \citet{1969mech.book.....L}. }. The result is that energy can
be quickly taken out of the parent mode and transferred to the
daughter modes, providing a means to limit the parent mode's
amplitude. Furthermore, the growth rate of the daughters is larger than
the growth rate of the parent
so that the daughters can ``catch up" to the parent even if they start from
a lower amplitude.

The parametric instability can be derived in the approximation where
the parent mode's amplitude is much larger than the daughter modes'
amplitudes so that the influence of the daughters on the parent
can be neglected. Performing a linear stability analysis of
eq.\ref{eq:3ampeqn}
\citep{1969mech.book.....L}, one
finds that the daughters grow exponentially like $\exp(t/\tau)$ when
the parent mode amplitude $A_1$ exceeds the critical value given by
\be
A_1^2 & = & {{\bar \gamma}_2 {\bar \gamma}_3 \over  \kappa^2
\omega_2 \omega_3} 
\left[ 1 + \left( \frac{\delta \omega} 
{{\bar \gamma}_{2}+{\bar \gamma}_{3}}  \right)^2 \right],
\label{eq:parametric0}
\ee
where ${\bar \gamma}_{2,3} = \gamma_{2,3} + 1/\tau$.  
In particular, the threshold at which the instability first starts
to operate is given by eq. (\ref{eq:parametric0}) at $\tau=\infty$,
\be
A_1^2 & = & \frac{1}{ \kappa^2 Q_2 Q_3}
\left[ 1 + \left( \frac{\delta \omega}
{\gamma_{2}+\gamma_{3}}  \right)^2 \right]
%\begin{array}{c}
%\mbox{parametric} \\
%\mbox{threshold}
%\end{array}
\label{eq:parametric}
\ee
where $Q_2 = \omega_2 / \gamma_2$ and $Q_3 = \omega_3 / \gamma_3$ are
the quality factors of the daughter modes.
In the limit $\gamma_2, \gamma_3 \to 0$ of negligible damping of the
daughter modes, it is useful to consider in addition the threshold
above which the daughter mode's growth
rate will exceed that of the parent mode.  This is given
by eq. (\ref{eq:parametric0}) at $\tau = \gamma_1^{-1}$, $\gamma_2 =
\gamma_3 = 0$:
\be
A_1^2 & = & \frac{1}{ \kappa^2 \omega_2 \omega_3}
[ \gamma_1^2 + \delta \omega^2/4 ].
\label{eq:parametric1}
\ee
We give an example showing the parametric instability in
Fig. \ref{fig:exampleevolution}.

Once the parametric instability occurs, the daughter modes start to
grow rapidly. We now discuss the conditions under which the
subsequent evolution leads to a saturation of the parent mode in the
three mode system.

Setting the time derivatives in eq.\ref{eq:3ampeqn} to zero, 
one finds the equilibrium solution for the parent
\be
A_1^2 & = & \frac{1}{\kappa^2 Q_2 Q_3}
\left[ 1 + \left( \frac{\delta \omega}
{\gamma_{2}+\gamma_{3}-\gamma_1}  \right)^2 \right],
%\begin{array}{c}
%\mbox{equilib-} \\
%\mbox{rium}
%\end{array}
\label{eq:equilsoln}
\ee
and daughter mode energies $A^2_{2,3}/A^2_1 = Q_{2,3}/Q_1$, where
$Q=\omega/\gamma$ is the quality factor of the mode.
Naively, one expects
that energy transfer from the parent to the daughters
occurs only if the daughter modes have a lower energy than the parent,
implying a lower daughter mode quality factor. This expectation is verified
by a stability analysis \citep{2001ApJ...546..469W,2000PhRvL..84..622D}
which shows that the equilibrium
solution is stable only when $\gamma_2+\gamma_3 \simgreat \gamma_1$
is satisfied.

More precisely, there are three different regimes in the three
dimensional space of parameters $(\gamma_2+\gamma_3)/\gamma_1$,
$\gamma_2 / \gamma_3$ and $\delta \omega / \gamma_1$.
First, the equilibrium solution is linearly stable
to small perturbations if two conditions are met \citep{1980PhFl...23.1142W}:
(i) the ratio of damping to driving is sufficiently
large $\gamma_{2}+\gamma_{3} \simgreat \gamma_1$,
and (ii) the detuning is sufficiently large,
$\delta \omega \simgreat (\gamma_{2}+\gamma_{3})/2$.
Second, in the regime where $\gamma_{2}+\gamma_{3}
\simgreat \gamma_1$ but where the detuning is small,
$\delta \omega \simless (\gamma_{2}+\gamma_{3})/2$, the
the amplitudes and phase undergo limit cycles characterized
by bounded, quasiperiodic orbits, as shown by
Fig.\ 2 of \citet{2001ApJ...546..469W}.
Those limit cycle solutions have time averaged parent mode amplitudes
comparable to the equilibrium amplitude (\ref{eq:equilsoln}), so
the equilibrium amplitude still characterizes the motion.
Third, if the daughter mode damping
is insufficient, $\gamma_{2}+\gamma_{3} \simless \gamma_1$,
all three amplitudes rise without bound and the solution is
nonlinearly unstable \citep{2000PhRvL..84..622D}.
%has recently derived a more exact criterion
%for this {\it nonlinear} stability, which shows that the region of phase
%space leading to bounded motion expands exponentially as the daughter
%mode damping exceeds the parent mode driving.
For our purposes, any solution
which is nonlinearly stable can saturate the growth of the r-mode, so that
the effective stability criterion is
\be
\gamma_{2}+\gamma_{3} \simgreat \gamma_1.
\label{eq:stable}
\ee

In the regime where the equilibrium solution is stable, it acts like
an attractor, and the system tends to evolve into this equilibrium
after the daughter mode amplitudes become comparable to that of the
parent mode.  The example shown in Fig.\ \ref{fig:exampleevolution}
exhibits this behavior, even though the system is started well away
from equilibrium.  Note that the equilibrium parent mode amplitude
(\ref{eq:equilsoln}) is always approximately equal to the threshold amplitude
(\ref{eq:parametric}), in the regime (\ref{eq:stable}) where the
energy transfer is stable.

The parametric instability can provide a means for
saturating the r-mode amplitude.  Suppose that a daughter pair exists
which is parametrically unstable for a certain value $A_1$ of the
parent mode amplitude, and that no other daughter pairs are unstable
at that amplitude.  Then, if the resonance is sharp, it is plausible
that only the parent and two daughter modes are relevant, and if
the condition (\ref{eq:stable}) is satisfied so that
the transfer of energy is stable, then driving of the r-mode by
gravitational radiation reaction can be balanced by nonlinear energy
transfer to the pair of daughter modes.
Thus, {\it the daughter mode pair for which the instability threshold
(\ref{eq:parametric}) is lowest sets the saturation amplitude for the
r-mode, if the stability constraint (\ref{eq:stable}) is satisfied for
that daughter mode pair}.  Daughter
pairs with higher thresholds will not be excited because the parent's
amplitude cannot rise much above the lowest threshold (see Fig.\
\ref{fig:exampleevolution}).

The task of finding the saturation amplitude in the weak driving regime
involves searching through all possible daughter mode pairs to minimize
the parametric threshold (\ref{eq:parametric}).
This amounts to
maximizing $\kappa$ while minimizing the mismatch
$\delta \omega^2 + (\gamma_{2}+\gamma_{3})^2$ subject to the
stability constraint. Once this ``best" daughter mode pair has been found,
the saturation amplitude is
\be
A_1^2 & \simeq  & \frac{1}{\kappa^2 Q_{2} Q_{3} } |_{\rm best\ pair},
\label{eq:Abest}
\ee
assuming that the strong resonance condition
$\delta \omega \simless \gamma_2 + \gamma_3$
is satisfied.  {\it We can state the following rule of thumb: for
coupling coefficients
of order unity, the r-mode will saturate to an amplitude less
than unity if the best daughter pair are high Q oscillators.} Quality
factors of low order global modes in neutron stars can easily be
$10^6$ or larger.

Finding the saturation amplitude in the weak driving regime
has now been reduced to the following physics problem.
First determine the oscillation modes present in the star. Calculate their
damping and driving rates, as well as the nonlinear coupling coefficients
between daughter pairs and the r-mode. Once the magnitude and scalings of
these quantities are known, reliable estimates of the parametric threshold
can be made (see Sec.\ \ref{sec:satd} below).

Finally, note that nonlinear coupling terms such as $ \dot{q}_2
\simeq \kappa_{211} q_1^2$
which couple the parent mode twice with a daughter 
mode have been ignored in eq.\ref{eq:3ampeqn}. Since these terms scale
as $A_1^2$, instead of $A_1$ as for the parametrically driven modes,
they are smaller in the weakly nonlinear regime.
In addition, the coupling coefficients drop off rapidly for this type of
coupling as the wavenumber of mode $2$ is increased (see appendix
\ref{app:kappa}). Hence, only daughter modes with comparable wavenumber
to the parent couple well. However, the resonance condition cannot be
finely tuned for comparable wavenumber modes, since there are so few of them.
As opposed to the couplings 
$\kappa_{211} q_1^2$, parametric type couplings
have the double advantage of allowing coupling of the parent mode with
daughter modes of arbitrarily large wavenumber, and the resonance condition
becomes satisfied to a higher degree of precision for large daughter mode
wavenumber.

\subsection{ the continuum limit }
\label{sec:continuum}

%
% EEF
%
% I modified the following figure and figure caption to more
% accurately reflect the flow of energy in phase space, and the fact
% that when an inertial range exists most of the energy is ultimlately
% dissipated by shear viscosity rather than bulk viscosity
%
\begin{figure*}
\centerline{\psfig{figure=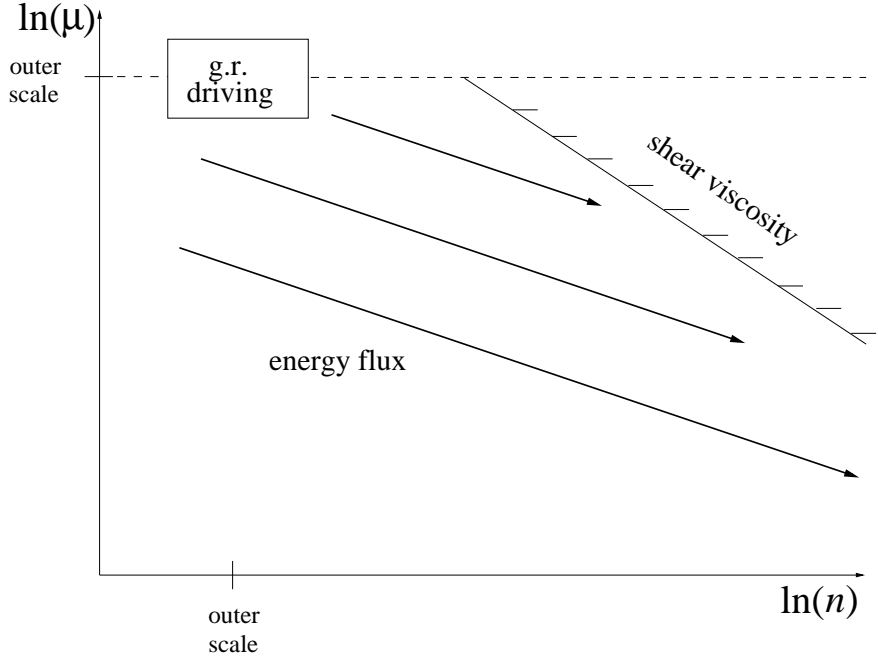,width=0.7\hsize}}
\caption[]{ Saturation in the continuum limit.
Energy is input to the r-mode at the outer scale of both wavenumber
$n$ and frequency $\mu$. 
Energy then cascades to small $\mu$ and large $n$ along the integral
curves $n^{\alpha_\mu} \mu^{\alpha_n} = $ (const) of the energy flux
vector field (\ref{eq:calF}).
% Eventually the energy reaches the region $n
%\mu^{3/2} \ge $ (const) or inner scale [cf.\ Eq.\
%(\ref{eq:shearregime})] where it is damped by shear viscosity. 
}
\label{fig:cascade}
\end{figure*} 

In the above ``discrete" scenario, the saturation amplitude of the
driven mode scales as $A \sim Q_d^{-1}$, where $Q_d$ is the quality factor
of a damped mode. In the ``continuum" picture that we now discuss,
the saturation
amplitude is {\it independent} of the linear damping rates, since the energy
is transferred by nonlinear interactions. In this cascade picture, both
the shape and normalization of the wave energy spectrum are set only by the
detailed nonlinear interaction between waves, and the energy input to the
system. 

How does the cascade arise? Imagine starting with a system in the weak driving
limit and adiabatically increasing the driving. When $\gamma_{gr}$ becomes
greater than the damping rate $\gamma_2+\gamma_3$ of the daughter pair with
the lowest threshold, the equilibrium solution for that pair is no longer
stable and the energy of all three modes will begin to grow. When the energy
has grown to the point that additional parametric thresholds are crossed,
and if the energy transfer to these pairs is stable, the driven mode will again be
saturated. As the driving is increased, this process will continue
until many daughter modes are excited with large amplitude.
Since linear damping is smaller than driving over a certain range of daughter
mode lengthscales, an \emph{inertial range}
% (not to be confused with inertial 
%oscillation mode) 
has formed where nonlinear forces are dominant. We 
now proceed to give a heuristic derivation of instability saturation 
in this continuum case, leaving the detailed derivation 
appendix \ref{app:cascade}.

Since many modes are excited, we treat the quantum
numbers for each mode as a continuum. Introducing the ``occupation number"
\footnote{ To get a quantity with the units of action, multiply $N_\alpha$
by $\Eunit$.}
(quasi-particle number) for mode $\alpha$
\be
N_\alpha=|q_\alpha|^2/|\omega_\alpha| \geq 0,
\ee
the mode energy becomes
\be
E_\alpha & = & 2\Eunit |q_\alpha|^2 = 2\Eunit |\omega_\alpha| N_\alpha.
\ee
Eq.\ref{eq:amplitudeeqn} describes both the fast variation of each individual
mode, as well as the slow variation due to nonlinear interactions 
between modes. We may average over the fast oscillations using the random
phase approximation \citep{1992kstw.book.....Z,1989ApJ...342..558K,
1998PhDT........28W} if the phase
randomization time set by the wave dispersion is shorter than
the nonlinear interaction timescale. Since the dispersion time
is comparable to the                           
mode period for inertial waves, this is equivalent to the weak nonlinearity
condition. The resultant kinetic equation for
the wave amplitudes is \citep{1992kstw.book.....Z,1989ApJ...342..558K}
\be
\dot{N}_\alpha & = & I_\alpha + \Gamma_\alpha N_\alpha
\label{eq:ndot}
\ee
where $I_\alpha$ represents the rate of change of $N_\alpha$ due to nonlinear
interactions, and $\Gamma_\alpha$ is the rate of driving ($>0$) or damping
($<0$). The interaction term has the form
\be
I_\alpha & = & s_\alpha \pi \sum_{\beta \gamma}
|\kappa_{\alpha \beta \gamma}|^2
|\omega_\alpha \omega_\beta \omega_\gamma|
\delta( \delta\omega_{\alpha \beta \gamma})
\nonumber \\ & \times &
\left( s_\alpha N_\beta N_\gamma
+ s_\beta N_\gamma N_\alpha + s_\gamma N_\alpha N_\beta \right)
\label{eq:kinetic}
\ee
where $s_\alpha$ is the sign of the frequency of mode $\alpha$.

To proceed further, we must introduce a few properties of the oscillation modes
to be derived in section \ref{sec:whichmodes}. Let
$n$, $k$, and $m$ denote the perpendicular (to $\Omegavec$), parallel,
and azimuthal number of nodes, respectively. Since inertial
mode oscillation frequencies are proportional to the rotation frequency, 
we write the mode frequency as $\omega=2\Omega\mu$, where $\mu^2 \leq 1$
is the dimensionless frequency.

Approximate stationary solutions of eq.\ref{eq:ndot}
are found in two steps (see \citet{1992kstw.book.....Z} for detailed
derivations). First, one ignores the
driving and damping, so that the energy flux is conserved.
In this case, the energy flux $\calF$ is defined by
\be
\omega \int dm\, I & \equiv & - \grad_k \cdot \calF
\ee
where $\grad_k$ is the gradient in momentum space and
we have integrated over the $m$ quantum number.
In appendix \ref{app:cascade}, we 
show that stellar inertial waves support a flux of energy to large $n$
and small $\mu$. 
A schematic drawing of this cascade is given in fig.\ref{fig:cascade}.
The occupation number \footnote{The scaling of this expression for the
occupation number can be simply derived from dimensional analysis
together with the fact that 3-mode interactions dominate over 4-mode
and higher order interactions. See, e.g.,
Sec.\ 3.3.1 of \citet{1992kstw.book.....Z}. }
for each mode is
\be
N & = & \Omega^{-1} N_0 n^{-4} |\mu|^{-1/2}
\propto n^{-7/2} |k|^{-1/2}
\ee
where the normalization constant is related to the fluxes $\calF^n$ and
$\calF^\mu$ in the $n$ and $\mu$ directions, respectively, by
\be
\calF^n & = & 8 \alpha_n N_0^2 n^{-1} |\mu|^{-1} \Omega \Eunit
\nonumber \\
\calF^\mu & = & - 8 \alpha_\mu N_0^2 n^{-2} \frac{\mu}{|\mu|} \Omega \Eunit.
\label{eq:calF}
\ee 
The constants $\alpha_n$ and $\alpha_\mu$ are order unity and positive.
A surface of constant energy in momentum space has $\mu \propto n^8$, showing
that the energy cascades to small frequencies quite rapidly with wavenumber,
because of the strong dependence of the coupling coefficients on frequency
(see appendix \ref{app:kappa}).

The final step is to match the inertial range solution to the driving range.
In other words, we need to find steady state solutions to the equation
$I + \gamma_{gr}N=0$, where $\gamma_{gr}$ is the driving rate by gravitational
radiation. We approximate $N$ in the driving region by extending the inertial
range solution.
The power input by the instability is given by 
\footnote{ Even if a relatively narrow region $\sim \Delta n^3$ 
in phase space is being
driven, \citet{1992kstw.book.....Z}
find that one should use the whole volume $\sim n_d^3$
instead of $\Delta n^3$ since a peak develops in the driving region.
Since the r-mode has relatively small wavenumber, the width of the driving
region occupied by the r-mode may be considered relatively wide ($\Delta n/n
\sim 1/3$, $\delta k/k \sim 1$, $\delta m/m \sim 1/2$).  
 }
\be
&& \mbox{input power} = \int dn\, dk\, dm\, \gamma_{gr} E 
\nonumber \\ & \simeq & \frac{2}{3} n_r^3 \gamma_{gr} E_r
= \frac{8}{3} \gamma_{gr} \Eunit N_0 n_r^{-1} \mu_r^{1/2}
\label{eq:inputpower}
\ee 
where the $r$ subscript denotes the driven r-mode and we have approximated
the (dimensionless) volume in phase space being driven as $ \propto n_r^3$.
The energy escaping the driving region is given by integrating up the flux
through each boundary.
Roughly, this is given by
\be
\mbox{output power} & = & \int_{\mu_r (1-1/n_r)}^{\mu_r} d\mu\ n\calF^n 
\nonumber \\ & + & 
\int_{n_r}^{n_r+1} dn\ n |\calF^\mu|
\nonumber \\ & \simeq & 
8 n_r^{-1} N_0^2 (\alpha_n + \alpha_\mu) \Omega \Eunit.
\ee
Equating the input to output power we can solve for the normalization 
constant 
\be
N_0 & = & \frac{1}{3} \frac{1}{\alpha_n + \alpha_\mu} \frac{\gamma_{gr}}{\Omega}
\mu_r^{1/2}.
\ee
Given the normalization, we find
\be
\frac{E_r}{\Eunit} & = & \frac{E(r-mode)}{0.5M\rstar^2\Omega^2} 
\nonumber \\ & = & 
\frac{2n_r^{-4} \mu_r}{3(\alpha_n + \alpha_\mu)} 
\frac{\gamma_{gr}}{\Omega}
\equiv \alpha_e \frac{\gamma_{gr}}{\Omega}
\label{eq:saturationenergy1}
\ee
where we parametrize our inexact treatment of the matching condition
with the parameter $\alpha_e=2n_r^{-4} \mu_r/3(\alpha_n + \alpha_\mu)$.
If we use the quantum numbers of the r-mode, $n_r=3$ and $\mu_r=1/3$,
we find $\alpha_e \simeq 4 \times 10^{-4}$. However, we choose to be
very cautious about this factor since we are extrapolating a WKB treatment
into the regime of low order modes \footnote{ The detuning may become
non-negligible for low order modes. In addition, the resonance width 
from $\gamma_{\rm gr}$ can become important for coupling directly to the 
unstable r-mode. }.
A more conservative estimate would
be to set $n_r=\mu_r=1$, giving $\alpha_e \simeq 0.1$. We will use the 
more conservative result for numerical work in the rest of this paper, 
but recall that it may {\it overestimate} the saturation amplitude by
up to three orders of magnitude. Using the r-mode driving rate from eq.
\ref{eq:gammagr}, the final result for the saturation energy is
\be
\frac{E(r-mode)}{0.5M\rstar^2\Omega^2}
& = &
10^{-6} \left( \frac{\alpha_e}{0.1} \right) \nu_{\rm khz}^5
\label{eq:continuumsat}
\ee
where $\nu_{\rm khz}$ is the spin frequency in units of $1000 \,\rm Hz$.

Why is the saturation amplitude so small? The factor $\gamma_{gr}/\Omega$
is inevitable
\footnote{ If the energy transfer is local in frequency space, this scaling
will also hold for interaction with other wave families, such as
inertial-gravity modes.}
since the only quantity with the units of frequency
in the nonlinear interaction rate is $\Omega$. The numerical factor
$\alpha_e$ depends on considerations such as the effective volume and
area of the driving region, and the power in the driving region relative
to the largest scale (energy bearing) waves.

Eq.\ref{eq:continuumsat} is one of the central results of this paper.
It applies when
nonlinear energy transfer is faster than linear damping.
If nonlinear energy transfer becomes slower than linear damping, the 
discrete limit of section \ref{sec:discrete} is recovered.
Note that the saturation amplitude decreases very rapidly with stellar
spin frequency.

% -----------------------------------------------------------
% -----------------------------------------------------------

\section{Oscillation modes in rapidly rotating neutron stars }
\label{sec:whichmodes}

In this section we discuss the oscillation modes present in 
rapidly rotating neutron stars.
We argue that at the rapid rotation rates of interest
for the r-mode instability, the buoyancy and elastic restoring forces
can be ignored in comparison with the Coriolis force.
The resulting modes which are restored by 
the Coriolis force are called {\it inertial modes}, of which the r-modes
are a subset.

\subsection{ motivation for inertial waves }

Within a minute after their birth in a supernova, neutron star cores 
have become transparent to neutrinos and cooled down sufficiently to
form a degenerate gas of neutrons, with a small admixture of electrons 
and protons determined by beta equilibrium. As shown clearly by
\cite{1992ApJ...395..240R},
the varying electron fraction $y_e \simeq 6.0 \times 10^{-3} (\rho/
2.8 \times 10^{14}\rm\ g\ cm^{-3})$ in the star causes a stable stratification
and resulting buoyancy force:
Since the neutron pressure
$p_n \propto [\rho/(1+y_e)]^{5/3}$,
displacing a fluid element upward on timescales slower than the sound
crossing time and faster than the timescale of the 
beta reactions results in the fluid element being
heavier than its surroundings, since it came from a region of larger $y_e$.
An oscillatory motion results, with maximum frequency of order the 
Brunt-Vaisala frequency \citep{1992ApJ...395..240R}
$N_{bv} \simeq (0.5 y_e g/H_p)^{1/2} \simeq 500\,
{\rm sec}^{-1}
\sim 2\pi \times 100\,\rm Hz $, where $g$ is the local gravity
and $H_p$ is the local pressure scale height.

The buoyancy force on a fluid element is just $F_{\rm buoy} =-g \delta \rho
\simeq - \rho N_{bv}^2 \xi^r$,
where $\xi^r$ is the radial component of the Lagrangian fluid
displacement. The Coriolis force is given
by $F_{\rm cor} \sim 2 \rho \Omega \omega \xi$, so that the ratio of these two
forces for $\omega \sim \Omega$ is roughly
\be
\frac{F_{\rm buoy}}{F_{\rm cor}} & \sim & \left( \frac{N_{bv}}{\Omega} \right)^2
\sim \left( \frac{100\,\rm Hz}{\nu_{\rm spin}} \right)^2.
\ee

In a detailed study of the solutions to the fluid perturbation equations
for rotating stars including buoyancy, \citet{YL2000} showed that in the
limit of ${F_{\rm buoy}}/{F_{\rm cor}} \gg 1$ the solutions are 
approximated very well by the r- and g-modes. This limit of large
buoyancy force was examined by \citet{morsink2002} who showed that 
the nonlinear couplings between r-modes are too small to cause
saturation to occur. In the limit of 
${F_{\rm buoy}}/{F_{\rm cor}} \ll 1$ \citet{YL2000} have shown
that the solutions of the perturbation equations are well approximated
by the inertial modes. As long as we restrict our calculations to 
stars spinning at a frequency greater than 100 Hz, the inertial 
modes with zero buoyancy are very good approximation. As we are
interested in the possibility of mode saturation at spin frequencies
at least as large as 300 Hz, the inertial modes are the most 
relevant modes and it is possible for us to ignore the buoyancy
force as a first approximation.
This enables us to find simple solutions for the modes if
we further approximate the shape of the star as spherical,
a valid approximation for rotation rates well below the breakup rate.
However, we expect that the qualitative results found here will hold even
in the case when buoyancy is included. The reason is that the
approximations made still provide a dense spectrum
of modes that may be arbitrarily resonant with the
r-mode in the continuum limit. Although the numerical value of the coupling
coefficients and damping rates may change because we don't have
exactly the correct shape of the 
eigenfunctions, we are confident that the essential qualitative features 
present in our simple example will carry over.

\citet{2001MNRAS.324..917L} have shown that the elastic restoring force
in the neutron star crust becomes small compared to the Coriolis force
above a rotation rate of $\sim 50 \,\rm Hz$. The net result is that core modes
can penetrate into the crust, with only a small discontinuity at the 
crust-core boundary because of the impedance mismatch. We will ignore crustal
elasticity for the remainder of this paper.

We have not included superfluid effects in our calculations. The principal
new effect would be dissipation due to mutual friction (the modes themselves
are not expected to be changed very much; see, e.g., \citet{2000PhRvD..61j4003L}).
However, we note that our estimate of the saturation amplitude does not 
depend on the dissipation rate if an inertial cascade forms (although the
outer scale of the inertial range might be affected). Thus, if saturation
involves a cascade of energy to numerous inertial modes, we still expect
our estimates to hold. Our estimates would change if dissipation via mutual
friction is strong enough that only a few modes are excited parametrically. 
We postpone a thorough examination of this case, which would depend on 
uncertain mutual friction coefficients, for another paper. However, either
way, {\it the saturation amplitude will still be very small}.

In the next subsection, we discuss inertial mode eigenfunctions
in weakly stratified stars.

\subsection{ stellar inertial modes }
\label{sec:inertialmodes}

We solve the Euler and continuity equations for adiabatic perturbations
of a rotating star. The background star is taken to be spherically symmetric
with uniform rotation rate $\Omegavec=\Omega \unit_z$ 
and negligible stable stratification.
Perturbation modes of the form $\exp(i m \phi - i\omega t )$ are found using
the Cowling approximation.

The Euler, continuity, and state equations are
\be
\rho \ddot{\xivec} + 2\rho \Omegavec \times \dot{\xivec}
& = & - \grad \delta p + \gvec \delta \rho
\label{eq:euler}
\ee
\be
\delta \rho + \grad \cdot \left( \rho \xivec \right) & = & 0
\label{eq:cont}
\ee
\be
\delta p & = & c^2 \delta \rho,
\label{eq:eos}
\ee
where we have ignored terms involving the Brunt-Vaisala frequency
\be
N_{bv} & = & \left( g \frac{ d \ln(p^{1/\Gamma_1}/\rho) }{dr} \right)^{1/2},
\ee
a valid assumption for $\omega \gg N_{bv}$ and $\Omega \gg N_{bv}$. 
Here we have defined the adiabatic index $\Gamma_1$.
In this limit,
the adiabatic sound speed $c$ and density scale height
$H$ are related by $c \simeq (gH)^{1/2}$.
Substituting the assumed dependence on $\phi$ and $t$, and eliminating
$\delta \rho$, we find that eqns. \ref{eq:euler} -- \ref{eq:eos} become
\be
\xivec + i q\ \unit_z \times \xivec & = & \grad \psi
\label{eq:euler2} \\
\grad \cdot \xivec + \frac{\omega^2}{c^2} \psi & = & \frac{\xi^r}{H}.
\label{eq:cont2}
\ee
Here we have replaced the Eulerian pressure perturbation by the quantity
$\psi$ defined by
$ \delta p = \rho \omega^2 \psi,$
and defined the dimensionless inverse frequency
$q=2\Omega/\omega$. We will also heavily use the dimensionless frequency
$\mu=1/q=\omega/2\Omega$.
We drop the term $\omega^2 \psi / c^2 \propto \Omega^2/(GM/\rstar^3)$
since we are working to leading order in $\Omega$; this is consistent
with our assumption that the background star is spherical and suffices
to compute the mode functions to leading order in $\Omega$.

The Euler equation \ref{eq:euler2} can be solved
\footnote{ The determinant of this
transformation is singular only if $\omega^2=4\Omega^2$.}
for $\xivec$ in terms of $\psi$:
\be
\xivec & = & (1-q^2)^{-1}
\left[ \grad \psi - q^2 \unit_z (\unit_z \cdot \grad \psi)
- iq \unit_z \times \grad \psi \right].
\label{eq:xi_vs_psi}
\ee
Substituting eq. \ref{eq:xi_vs_psi} into the continuity equation 
\ref{eq:cont2} gives the wave equation 
\be
\grad^2 \psi - q^2 \frac{\partial^2 \psi}{\partial z^2}
& = & 
H^{-1} \left( \frac{\partial \psi}{\partial r} 
- q^2 \cos\theta \frac{\partial \psi}{\partial z} - \frac{mq}{r} \psi  \right).
\label{eq:wave_eqn}
\ee

The boundary condition near the surface is that the Lagrangian change
in the pressure vanish, $\Delta p = \delta p + \xivec\cdot\grad p =0$,
so that 
$\delta p \simeq \rho g \xi^r$. 
%Near the surface, the boundary condition is that each term in eq.\ref{eq:cont2}
%must remain finite. This implies that the Lagrangian pressure perturbation
%$\Delta p = \delta p + \xivec\cdot\grad p$ must go to zero as fast as the
%pressure, or equivalently, that $\delta p \simeq \rho g \xi^r$.
This is just the statement
that the surface layer is hydrostatic, a consequence of the vanishing
sound crossing time across a scale height for small depth.

Equation \ref{eq:wave_eqn} does not appear to be solvable by separation of
variables\footnote{By separation of variables, we mean that (1) the differential
equation is separable, and (2) the boundary conditions are applied on a 
surface where one of the coordinates is constant.}.
This motivates us to 
examine approximate solutions valid for short wavelengths.
Our solution generalizes the exact solution of
\citet{1889PTRSL...A180..187} for the constant density
star; in fact our solution is just Bryan's
solution divided by $\sqrt{\rho}$.

Defining
\be
\psi(\xvec) & = & \psi_0 \left( \frac{\rho_0}{\rho} \right)^{1/2} f(\xvec)
\label{eq:definef}
\ee
where $\psi_0$ is a normalization constant and $\rho_0$ is the central density,
the differential equation
for $f$ is
\be
\grad^2 f - q^2 \frac{\partial^2 f}{\partial z^2} + \calK^2 f & = & 0
\label{eq:newbryan}
\ee
with
\be
\calK^2 & = & 
 (1-q^2 \cos^2\theta) \left( \frac{1}{2} \frac{dH^{-1}}{dr} - \frac{1}{4H^2}
 \right) 
\nonumber \\ & + & 
 \frac{mq}{rH} + \frac{1}{2rH} \left( 2 - q^2 \sin^2\theta \right). 
\label{eq:calK2}
\ee
The first two terms in eq. \ref{eq:newbryan} are just the usual differential
equation for inertial modes, as derived by Bryan.
The compression term $\xi^r/H $ 
in the continuity equation \ref{eq:cont2} is imaginary 
in the WKB limit, and leads to the WKB envelope $\rho^{-1/2}$.
The definition in eq.\ref{eq:definef} accounts for this envelope, so that
the correction terms in eq.\ref{eq:calK2} are now real.
The $\calK^2 f$ term is most important near the surface,
where it scales as $(|\kvec|H)^{-2}$ relative to the other terms
in eq.\ref{eq:newbryan} ($|\kvec|$ is the local WKB wavenumber).
This term describes
a slow variation of the wavenumber with position, and is negligible
for short wavelength modes. Henceforth, we set $\calK \simeq 0$ in the 
interior of the star.

The short wavelength approximation breaks down
when the vertical wavenumber $k_r$
\footnote{ One must use care evaluating $k_r$ for inertial modes near
the surface of the star, since it can vary strongly with the angle $\theta$.
Qualitatively, this strong variation occurs because one is imposing a 
spherical boundary condition on waves with inherent cylindrical symmetry.}
is comparable to the scale height, at
about $2k_rH \sim 1$, as can be verified directly from eq. \ref{eq:newbryan}.
When the eigenfunction is constant
over a scale height one may picture that the wave attempts to lift an entire 
scale height of material, which causes reflection.
We extend our interior solution to the surface by replacing $\rho$ in 
eq. \ref{eq:definef} with a ``cutoff" value $\rho_{cut}$
which becomes constant about 
one wavelength from the surface. \footnote{ If the density profile near 
the surface is a power law with depth, one can separate variables in 
the bi-spheroidal coordinates introduced below. These more rigorous 
solutions close to the surface agree 
with the cutoff behavior described here for the density. Although one
could, in principle, match the interior WKB solution to the surface solution,
the cutoff for the density gives an adequate approximation for the problem
at hand.}

\begin{figure*}
%\centerline{\psfig{figure=,width=0.45\hsize}}
\begin{picture}(200,200)(0,0)
\put(50,0){\psfig{figure=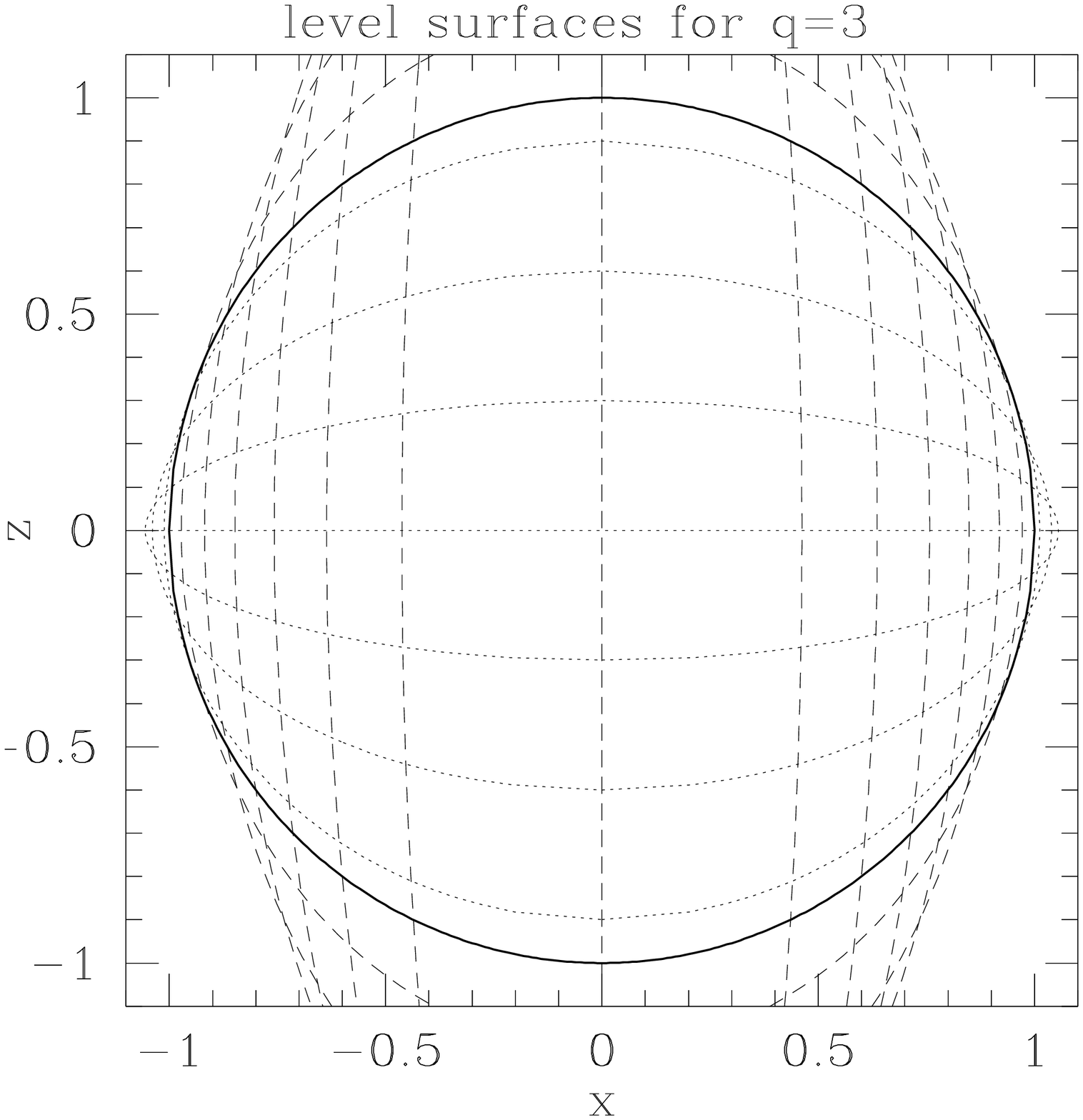,width=0.35\hsize} }
\put(250,0){ \psfig{figure=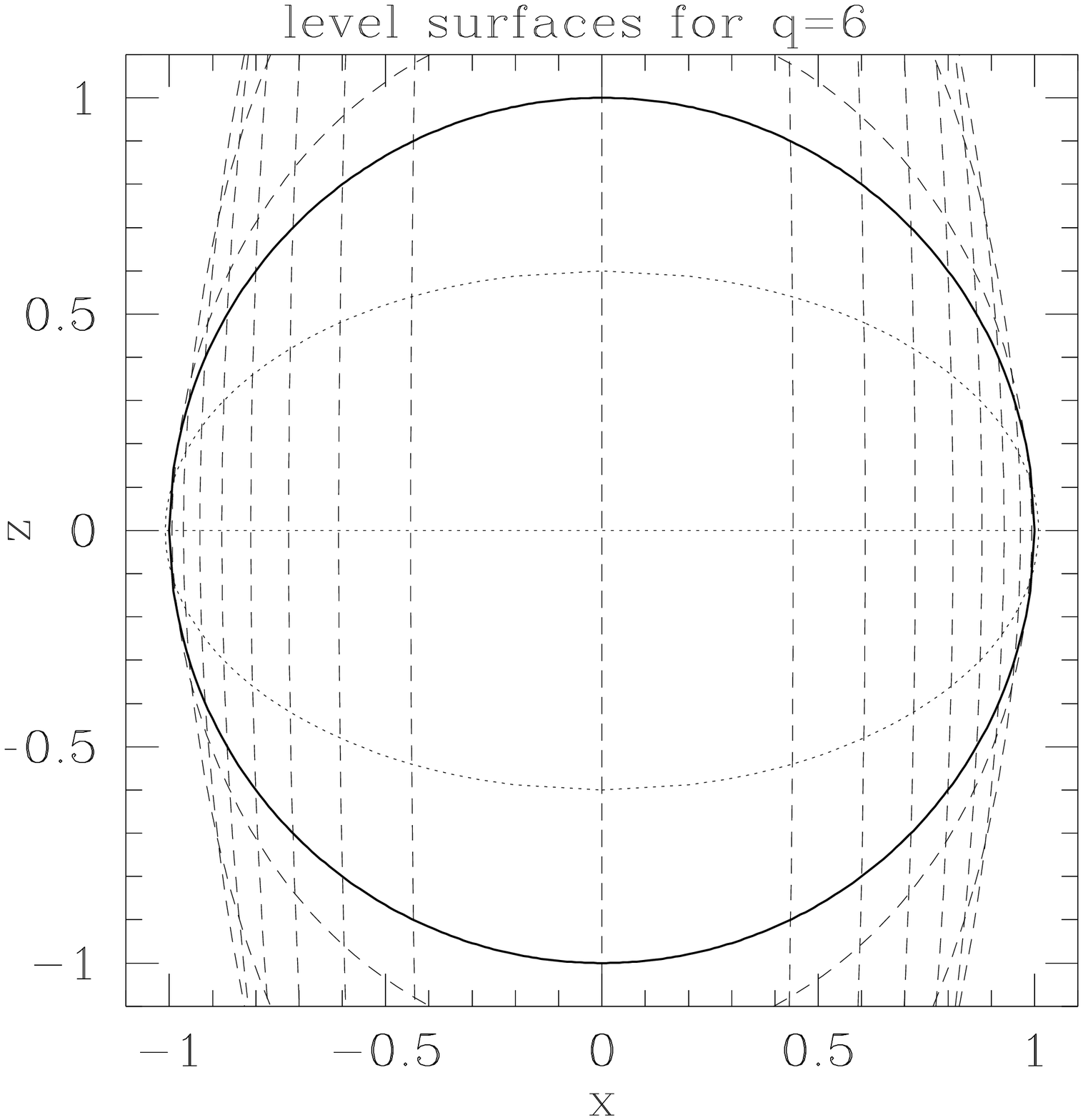,width=0.35\hsize} }
%\put(0,0){ \psfig{figure=level_surf_1.1.ps,width=0.35\hsize} }
%\put(200,0){ \psfig{figure=level_surf_10.ps,width=0.35\hsize} }
\end{picture}
\caption[]{
Surfaces of constant bi-spheroidal coordinate for
different values of $q$. The surface of the star is the thick solid line
on the unit circle. Only the portions of the spheroids inside this
circle are relevant. The short dashed lines represent surfaces of constant
bi-spheroidal coordinate $-1/|q| \leq x_2 \leq 1/|q|$ while the long
dashed lines are for the second bi-spheroidal coordinate, which takes on the
range $1/|q| \leq x_1 \leq 1$.
The level surfaces are at the values $0.1,0.2,...,0.9$.
The surface of the star is described by two coordinate patches, $x_1 =
1/|q|$ near the equator, and $x_2 = 1/q$ near the pole.
}
\label{fig:level_surf}
\end{figure*}

\citet{1889PTRSL...A180..187}
found a solution to eq.\ref{eq:newbryan} for $\calK=0$
in terms of an ingenious bi-spheroidal
coordinate system that depends on the frequency $\mu$.
\citet{1999PhRvD..59d4009L} have given a careful discussion of
this coordinate system, paying particular attention to the behavior
of the coordinates at the surface of the star. 
Define the bi-spheroidal coordinates 
$|\mu| \leq x_1 \leq 1$ and $-|\mu| \leq x_2 \leq |\mu|$ by
\be
x & = & 
\rstar \left[ \frac{(1-x_1^2)(1-x_2^2)}{1-\mu^2} \right]^{1/2} \cos \phi
\nonumber \\
y & = & \rstar
 \left[ \frac{(1-x_1^2)(1-x_2^2)}{1-\mu^2} \right]^{1/2} \sin \phi
\nonumber \\
z & = & \rstar \frac{x_1 x_2}{|\mu|}.
\label{eq:ellipoidalcoords}
\ee
In fig. \ref{fig:level_surf} 
we plot the surfaces of constant bi-spheroidal
coordinate in the $x-z$ plane. In either the $\mu \sim 1$ or $\mu \ll 1$
limits, the surfaces of constant $(x_1,x_2)$ are nearly in the 
$z$ and $R$ directions over most of the star, 
as one would expect for a local plane wave propagating in the $z$
or $R$ direction. (Here $R$ is the cylindrical radius.)
However, the coordinate lines near the point
$r \sim \rstar$ and $\cos\theta = \pm |\mu|$ on the surface of the star
vary quite rapidly with respect to $r$ and $\theta$. The
result is that the WKB wavenumber becomes quite large near these
singular points. (We give a detailed mathematical discussion in 
appendix \ref{app:kappa}.) As one can see from fig. \ref{fig:level_surf},
the coordinate lines come closer to the surface near the singular
points, implying the upper turning point is much closer to the surface
near the equator (for small $\mu$) than the poles. As a result, the wave
amplitudes will be much larger near the equator, as we will now show. 

Our approximate solution for the interior of the star is to ignore terms
of order $(|\kvec|H)^{-2}$, so that the differential equation
becomes
\be
\grad^2 f - q^2 \frac{\partial^2 f}{\partial z^2} & = & 0.
\label{eq:bryan}
\ee
Changing to bi-spheroidal
coordinates in eq. \ref{eq:bryan} gives separable differential
equations (see \citet{1889PTRSL...A180..187} and
\citet{1999PhRvD..59d4009L} for details) .
Define the solution $f(x_1,x_2) = f_1(x_1)f_2(x_2)$
where both $f_1$ and $f_2$ satisfy
\be
\frac{\partial }{\partial x} \left[ (1-x^2)\frac{\partial f}{\partial x} \right]
+ \left( \kappa^2 - \frac{m^2}{1-x^2} \right)f & = & 0
\label{eq:legendre}
\ee
for separation constant $\kappa^2$. This equation has the Legendre
function solutions 
$\kappa^2=n(n+1)$, $f_1(x_1)=P_{nm}(x_1)$ and $f_2(x_2)=P_{nm}(x_2)$.

The resulting solution for $\psi$ is then
\be
\psi(\xvec,t) & = & \psi_0 \left( \frac{\rho_0}{\rho} \right)^{1/2}
P_{nm}(x_1) P_{nm}(x_2)
e^{im\phi - i \omega t}.
\label{eq:interior_soln}
\ee
Note the important fact that {\it this solution is valid for an arbitrary
density profile $\rho$, so long as one is safely in the short wavelength
limit}. This is true even when $\rho$ is {\it not} a separable function
of $x_1$ and $x_2$, as is generally the case in the interior since
$\rho(r)=\rho \left( 
\sqrt{1 - (x_1^2-\mu^2)(\mu^2-x_2^2)/\mu^2(1-\mu^2)} \right)$. 

The r-modes do not have short wavelengths and hence cannot be described by
the above WKB approximation. However, in the leading order approximation
of a spherical background star with no buoyancy,
the r-mode solutions are given by \citep{1889PTRSL...A180..187}
\be
\psi(\xvec,t) & =& \psi_0 P_{|m|+1,m}(x_1)P_{|m|+1,m}(x_2)\exp(im\phi-i\omega t)
\nonumber \\ & \propto &
z R^{|m|} \exp(im\phi-i\omega t)
\label{eq:rmode}
\ee
and have frequencies $\mu=-{\rm sign}(m)/(|m|+1)$.

% This formula is valid
%even for compressible stars since $\xi^r=0$. Note that the r-mode
%frequencies {\it must} decrease toward zero as $m$ increases. 
%Hence, the interaction of three r-modes can never become arbitrarily resonant
%by going to large daughter mode $m$. Furthermore, \citet{2002PhRvD..65b4001S}
%have shown that coupling of three r-modes as in eq.\ref{eq:rmode} has
%zero coupling. \citet{morsink2002} has investigated the coupling of three
%r-modes in stars with buoyancy. Although buoyancy changes the r-mode 
%eigenfunction to allow coupling, Morsink also found that the detuning did
%not become small in the continuum limit.
%This is why the more general class of inertial
%modes are important for nonlinear transfer of energy.

We now derive the WKB limit for the solution in eq. \ref{eq:interior_soln}.
Writing $x_{1,2}=\cos\theta_{1,2}$  and substituting $f \propto 
\sin^{|m|}\theta\, \exp(i \int^\theta d\theta\, k)$ in eq. \ref{eq:legendre},
we find the following standing wave solutions:
\be
f(\theta) & \simeq & \frac{\pi^{-1}}{\sin^{1/2}\theta} \cos(p\theta+\alpha)
\ee
where the wavenumber is given by
\be
p & = & \left( n(n+1) - |m|[|m|+1] \right)^{1/2} \simeq n
\ee
for $n \gg |m|$ (the WKB limit) and 
\be
\alpha & = & -p\pi/2 \mbox{ for even parity }
\nonumber \\
\alpha & = & -(p+1)\pi/2 \mbox{ for odd parity modes}.
\ee
We have chosen to normalize the Legendre polynomials to unity over $4\pi$
steradians.
Note that the nodes are spaced evenly in $\theta_{1,2}$.
This WKB approximation
to the Legendre equation fails within about one wavelength of $\theta=0,\pi$.
The $\sin \theta_{1,2}$
 factor causes an increase in amplitude toward
$\sin\theta_{1,2}=0$. The collected result is then 
\be
\psi(\xvec) & = & \frac{\psi_0}{\pi^2}
\frac{\cos(p\theta_1+\alpha) \cos(p\theta_2+\alpha)}
{\left( \frac{\rho}{\rho_0}
\sin\theta_1 \sin\theta_2 \right)^{1/2} } e^{im\phi}.
\label{eq:efunctions}
\ee
The factor in the denominator $\rho \sin\theta_1\sin\theta_2 \propto
\rho R$ is just the mass element, and enforces roughly equal kinetic
energy in between each pair of nodes.

An approximate dispersion relation is easily derived using the eigenfunctions
of eq.\ref{eq:efunctions}. The boundary condition is that the compression
term $\xi^r/H$ in eq. \ref{eq:cont2}
must remain finite as $H \rightarrow 0$, implying $\xi^r
\rightarrow 0$ in the low frequency approximation.
At either surface patch $x_1=|\mu|$ or 
$|x_2|=|\mu|$, this condition implies
\be
(1-x^2) \frac{dP_{nm}}{dx} + m P_{nm} & = & 0
\label{eq:boundarycondition}
\ee
at $x=\pm |\mu|$. Eq.\ref{eq:boundarycondition} is equivalent to the one given by
\citet{1889PTRSL...A180..187}, as noted by \citet{1999PhRvD..59d4009L}. The 
dependence of the frequency and wavenumber on the background stellar model,
as discussed by \citet{1999ApJ...521..764L},
is small in the WKB limit.
Substituting the WKB expressions gives
\be
\sin\theta \tan(p\theta+\alpha) & = & - \frac{m}{p}.
\ee
In the limit $p \gg |m|$, the solutions are found by inspection to be
$p\theta+\alpha=-k\pi$, for the mode index $k$.
Including the finite $m$ term to first order gives the solution
\be
\mu_{nkm} & = & \omega_{nkm}/2\Omega 
\nonumber \\ & = &  \sin(k\pi/p + m/p^2 - \pi\delta/2p) 
\label{eq:dispersionrelation} \\ & \simeq & 
 k\pi/n
\label{eq:dispersionrelation2}
\ee
where $\delta=0$ for even parity modes and $\delta=1$
for odd parity modes. The $m$ term can be dropped except for the very low
frequency, even parity mode with frequency $\mu \simeq m/p^2$. All
other modes have $\mu \propto p^{-1}$. We find the approximate formula
in eq.\ref{eq:dispersionrelation} to agree quite well with the exact 
solutions of eq.\ref{eq:boundarycondition} even for $p$ as small as 5.
Lastly, we note that \citet{1999ApJ...521..764L} have checked the 
eigenmodes found using bi-spheroidal coordinates with those from a numerical
code in spherical coordinates, finding agreement.

We choose to normalize the eigenfunctions so that at unit amplitude
($A=1$) all modes have the same energy, which we call $2\Eunit$.
We can analytically compute the mode energy in the WKB limit where the 
eigenfunctions are rapidly oscillating ($n \gg m$), with the result
\be
E & = & \omega^2 \int d^3x\, \rho |\xivec|^2
\nonumber \\ & \simeq &
 4\pi^{-2} \frac{p^2 \mu^4}{(1-\mu^2)^{3/2}} \rho_0 \Omega^2 r_* \psi_0^2.
\label{eq:modeenergy}
\ee
This formula agrees well with numerical integrations.
Our normalization convention is that at unit amplitude all modes have the
same energy $2\Eunit$. We then find the value of the normalization
constant
\be
\psi_0^2 & = & \frac{ \Eunit } { 2\pi^{-2} \frac{p^2 \mu^4}{(1-\mu^2)^{3/2}}
 \rho_0 \Omega^2 r_*} \propto p^{-2} \mu^{-4}.
\ee
Modes with rapid spatial variation ($p \gg 1$) or larger frequency $\mu$
have smaller normalization in order for the energy to be the same.
As $\mu \rightarrow 1$, the wave amplitude goes to zero since inertial modes
do not exist outside this range.

Before moving on to discuss the nonlinear force coefficients, we discuss
the normalization integral in eq.\ref{eq:modeenergy}. One can
easily find the mode energy to leading order in $\mu$ by setting $\mu=0$
in eq.\ref{eq:modeenergy}. In this limit, the bi-spheroidal coordinates become
$x_1 \simeq (1-R^2)^{1/2}$ and $x_2 \simeq 0$. 
In this limit, the integrand is constant in $z$, and varies as
$(1-R^2)^{-1/2}$ with $R$, which is large near the surface. The kinetic
energy then converges as $(1-R^2)^{1/2}$ from the surface.

% -------------------------------------------------------------

\section{ coupling coefficients }
\label{sec:coupling}

The lowest order nonlinear interaction couples three inertial waves,
implying quadratic nonlinear
terms as in eq.\ref{eq:amplitudeeqn}. The expressions for the nonlinear
force coefficients can be derived either from an action principle
\citep{newcomb1962,1989ApJ...342..558K,1996ApJ...466..946K} or directly
from the equation of motion \citep{2002PhRvD..65b4001S}.
Note that Schenk et.al. have stressed
that the {\it form} of the coupling coefficient is the same for 
rotating systems as for nonrotating systems; only the explicit expressions
for the eigenfunctions and background stellar model need be modified.
Since we are using daughter modes with wavelengths much smaller than a stellar
radius, we keep only the largest \footnote{ Inertial waves in an infinite
homogeneous, incompressible medium have a nonlinear coupling as given here.
Including terms arising from compressibility or variation of the background
stellar quantities then gives terms which are small in the limit 
$|\kvec|H \gg 1$.
We ignore these terms here for simplicity, 
although the r-mode is formally a large
lengthscale mode. }
term in the coupling coefficient in 
an expansion of $(|\kvec|H)^{-1}$. For modes $\xivec_1$, $\xivec_2$,
$\xivec_3$, the {\it dimensionless} coupling coefficient \footnote{
see Schenk et.al. for a derivation of eq.\ref{eq:amplitudeeqn} and
the explicit form of the dimensionless coupling coefficient $\kappa$}
is
\be
\kappa_{123} & \simeq & - \frac{1}{2\Eunit}
\int d^3x  \left( \xi^i_1 \xi^j_2 \delta p_{3;ij}
\right. \nonumber \\ & + & \left.
 \xi^i_2 \xi^j_3 \delta p_{1;ij}
+ \xi^i_3 \xi^j_1 \delta p_{2;ij}
\right).
\label{eq:kappa}
\ee
Since $\delta p \propto \Omega^2$, we find that the interaction energy,
$\kappa \Eunit$, scales as the rotational energy of the star.
A natural unit of energy is then $\Eunit=0.5M\rstar^2\Omega^2$.
In these units, $A^2$ is the mode energy
in units of $2\Eunit=M\rstar^2\Omega^2$.

In section \ref{sec:epconservation} we discuss conservation rules for the
nonlinear coupling coefficients. Effectively, these rules pick out the largest
possible coupling coefficients. The scalings for $\kappa$ are discussed
in section \ref{sec:kappaestimate}. We confirm a result found in previous
studies \citep{2001ApJ...546..469W} that {\it for waves which satisfy the 
conservation rules, the coupling coefficients do not become smaller 
as the daughter mode wavenumber is increased}; even though each 
individual eigenfunction is highly oscillatory, the product is relatively 
constant. Numerical results are presented in section \ref{sec:kappanumerical}
and a detailed analytic calculation is given in appendix \ref{app:kappa}.

\subsection{ energy and momentum conservation }
\label{sec:epconservation}

\begin{figure*}
\centerline{\psfig{figure=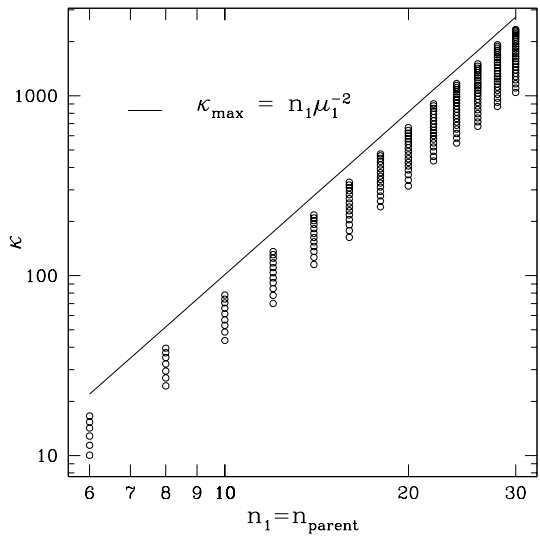,width=0.45\hsize}}
\caption[]{
Coupling coefficient as a function of parent mode quantum number $n_1$.
The other parent mode quantum numbers are held fixed at $k_1=1$ and $m_1=2$.
The daughter mode quantum numbers $n_2$ and $n_3$ are allowed to vary, but
we have fixed $k_2=-2$, $m_2=-3$, $k_3=-1$, and $m_3=1$.
The dots show the coupling coefficient determined
by numerical integration as described in the text.
The line gives the analytic approximation
$\kappa_{123} = n_1 \mu_1^{-2} \simeq \pi^{-2} n_1^3 k_1^{-2}$.
}
\label{fig:n1scaling}
\end{figure*}

Consider a parent mode with quantum numbers $(n_1,k_1,m_1)$ and frequency
$\mu_1$. 
We are free to choose daughter mode quantum numbers $(n_2,k_2,m_2)$
and $(n_3,k_3,m_3)$ in order to find the largest coupling coefficient
(see e.g. \citet{2001ApJ...546..469W}).
The integrand is highly oscillatory unless the phases of the waves match
at each point in the star. If we expand the standing wave solution 
in eq.\ref{eq:efunctions} in terms of travelling waves, a non-oscillatory 
integrand implies momentum conservation for the three travelling waves.
In addition to conservation of the $m$ quantum number, due to axisymmetry
of the background star, we also have momentum conservation along the 
$\theta_1$ and $\theta_2$ directions. For small $\mu$, the total number of
nodes along $\theta_1$ and $\theta_2$ simplifies to 
$N_1 \sim p \simeq n$ and $N_2 \sim |\mu|p/\pi \sim k$. The approximate
conservation laws which lead to large $\kappa$ can then be written
\be
m_1 + m_2 + m_3 = 0 \ \ \ \mbox{ angular momentum }
\nonumber \\
|n_2-n_3| \simless n_1 \ \ \ \mbox{ momentum along $\theta_1$ }
\nonumber \\
||k_2|-|k_3|| \simless |k_1| \ \ \ \mbox{ momentum along $\theta_2$ }.
\label{eq:conservation}
\ee
For small frequency, the $\theta_1$ and $\theta_2$ directions lie nearly
along the $R$ and $z$ directions, so that the second and third momentum
conservation rules correspond to conservation of momentum along $R$ and $z$.
In the limit that the daughter modes have much smaller wavelengths than the
parent mode, which will turn out to be the important limit, we find the 
simple result $n_2 \simeq n_3$ and $|k_2| \simeq |k_3|$; momentum conservation
implies that the daughter modes have momenta of equal magnitude and 
oppositely directed.

So far, we have used momentum conservation to determine three of the six
daughter mode quantum numbers. In order for energy to be efficiently 
transferred between modes, the interaction must be as nearly resonant 
as possible, meaning that the detuning is small:
\be
\delta \omega  = \omega_1 + \omega_2 + \omega_3 \simeq 0.  
\ee
There are two simple limits of interest. For short wavelength
daughter modes with $n_2 \simeq n_3$ and $k_2 \simeq k_3$, one 
has $\mu_2 \simeq \mu_3 \simeq - \mu_1/2$;
the parent mode interacts with nearly identical daughter modes of half
the frequency of the parent.
The second solution is where $n_1 \simeq n_2$, and $n_3 \ll n_1,n_2$.
In this case we find $n_2 \simeq n_1 + n_3$, $k_2 \simeq k_1$,
and $k_3 \simeq k_1(n_3/n_1)^2$. The frequencies are then $\mu_2 \simeq -\mu_1$
and $\mu_3 \simeq \mu_1(n_3/n_1) \ll \mu_1$.

\subsection{ analytic estimates }
\label{sec:kappaestimate}

Here we give a back of the envelope estimate for the coupling coefficient,
leaving the more detailed calculation for appendix \ref{app:kappa}.
We will only consider the important limit of short wavelength, nearly
identical daughter modes with $\mu_2 \simeq \mu_3 \simeq -\mu_1/2$.
We shall ignore factors of order unity for the present, concentrating
only on the scalings. As $\kappa$ is dimensionless, we set $\rstar=1$ in
this section for simplicity.

Incompressibility of the waves implies
\be
\kappa \sim \frac{1}{\Eunit} \int d^3x\, \rho \omega_1^2 k_{1z}^2
\psi_1 \xi_{2z}^2.
\label{eq:kappaapprox}
\ee
For the polytrope of index $1$ we find $\rho \propto \rstar-r \equiv
\zhat$ near the
surface, where $\zhat$ is the distance from the surface.
Since the WKB envelope of the waves rises steeply toward the surface, 
and the factor of $\rho$ cancels the $\rho^{-1}$ from $\xivec_{2z}^2$,
we find that the dominant contribution comes above the turning point for
the parent mode, where 
\be
\psi_1 & \sim & \frac{1}{ p_1\mu_1^2 \zhat_1^{1/2}} .
\ee
Here $\zhat_1$ is the turning point depth of the parent mode.
The daughter mode eigenfunction is strongly peaked in the $\theta$ 
direction due to the wavenumber
\be
k_{2z} \sim \frac{p_2 \mu_2 x_2 }
{ \left[ (\cos^2\theta - \mu_2^2)^2 + 8 \mu_2^2 \zhat \right]^{1/2} }
\ee
where $\theta$ is the polar angle in spherical coordinates.
The displacement for the daughter mode is then
\be
\xi_{2z}  & \sim & \zhat^{-1/2} 
\left[ (\cos^2\theta - \mu_2^2)^2 + 8 \mu_2^2 \zhat \right]^{-1/2}.
\ee
For $\cos\theta \simeq |\mu_2|$, $k_{1z} \simeq p_1$ since it is well
away from the singularity for mode 1 at $\cos\theta = |\mu_1|$.
Plugging into eq.\ref{eq:kappaapprox} we find
\be
\kappa \sim p_1 \zhat_1^{-1/2} \int_{\zhat_2}^{\zhat_1} d\zhat \int_{-1}^1
\frac{d(\cos\theta)}{  (\cos^2\theta - \mu_2^2)^2 + 8 \mu_2^2 \zhat }.
\ee
The integrand has a width $ d(\cos\theta) \sim \zhat^{1/2}$ and a height 
$(\mu_2^2 \zhat)^{-1}$, giving an area $(\mu_2^2 \zhat^{1/2})^{-1}$.
Using $|\mu_2| \sim |\mu_1|$ and $\zhat_1 \gg \zhat_2$, 
the final result is then
\be
\kappa & \sim  &\frac{ p_1}{\mu_1^2 \zhat_1^{1/2}}
\int_{\zhat_2}^{\zhat_1} \frac{d\zhat}{\zhat^{1/2}}
\simeq p_1 \mu_1^{-2}.
\label{eq:kapparesult}
\ee
The detailed calculation in appendix \ref{app:kappa} confirms that the
coefficient is about unity.

We now comment on the scalings for the {\it maximum} coupling coefficient
in eq.\ref{eq:kapparesult}. The maximum coupling coefficient is found
to be independent of the daughter mode quantum numbers. 
The reason, elucidated by \citet{2001ApJ...546..469W},
is that one is integrating over the daughter mode kinetic energy 
$\rho \mu_2^2 \xi_{2}^2$. This quantity is normalized to $2\Eunit$ when
integrated over the whole star, and is roughly $\Eunit \times \zhat_1^{1/2}$
when integrated over $0 \leq \zhat \leq \zhat_1$. Next, the factor $p_1
\sim n_1$ implies shorter wavelength parent modes interact more strongly.
This factor would appear for coupling of local waves in a box. However, 
the factor $\mu_1^{-2}$ would not appear for local waves in a box; it arises
from the large peak in the integrand near the surface. 

One might wonder whether or not the approximate conservation laws for 
colliding WKB waves will hold since one is integrating over a small
region of the star. The dominant 
contribution to the integrand comes from a region of size $\zhat \sim \zhat_1
\sim |\mu_1|/n_1$, and the angular size is $d(\cos\theta) \sim \zhat^{1/2}
\sim (|\mu_1|/n_1)^{1/2}$. The daughter modes have wavelengths $n_2$ or
$\mu_2 n_2$, depending on direction, so there are still sufficient 
oscillations in the important region of the star for large enough $n_2$.

\subsection{ numerical calculation }
\label{sec:kappanumerical}

We compute the integral in eq.\ref{eq:kappa} numerically as follows.
Choose a point in the star at which to evaluate the integrand. Evaluate
$\psi$ and $\delta p$ on the vertices of a Cartesian cube about this point.
The derivatives in eq.\ref{eq:xi_vs_psi} and \ref{eq:kappa}
can then be taken along Cartesian basis vectors
\footnote{ We evaluate vector quantities along Cartesian basis vectors
to avoid ``curvature terms" \citep{2001ApJ...546..469W} arising from
differentiating curvilinear basis vectors. Wu and Goldreich found these
terms are quite large, and cancel out in the end, so that significant
cancellation error can occur. We avoid such cancellation error by using
Cartesian basis vectors.}
, and then appropriate 
sums over indices taken. The resulting scalar integrand is independent
of the coordinate $\phi$ since $m_1+m_2+m_3=0$, so that only a two-dimensional
integral over $r$ and $\theta$ remains. We perform this integration with
second order accuracy, and increase the number of grid points until the
integral converges.

In fig.\ref{fig:n1scaling}, we show the numerical integrations for the
coupling coefficient as a function
of $n_1$, but fixed $k_1$ and $m_1$. We also fix $(k_2,m_2)$ and $(k_3,m_3)$
but allow $n_2$ and $n_3$ to vary. For a given $n_1$ we see there is a 
variation in $\kappa$ due to the degree of momentum conservation.
However, the upper envelope set by the {\it maximum} coupling coefficient
agrees to within $\sim 10\%$ of our analytic formula.

% -------------------------------------------------------------
% -------------------------------------------------------------

\section{ damping and driving rates }
\label{sec:damping}

We review the driving rate by gravitational radiation, and
derive simple analytic estimates for the damping rates of inertial modes.

% -------------------------------------------------------------

\subsection{ driving rate }

Gravitational radiation reaction is a driving force if the phase velocity 
in the azimuthal direction is positive in the inertial frame and negative
in the rotating frame; otherwise it damps the mode \citep{1998ApJ...502..714F}.
The driving rate falls off extremely rapidly with
wavenumber,
so that only the very lowest modes have an appreciable driving
rate compared to damping. \citet{1999ApJ...521..764L} 
have computed these driving
rates for the inertial modes of a polytrope of index 1,
and identified several low order driven modes. However, since the 
most unstable mode by far is the $(n,k,m)=(3,1,2)$ r-mode, we can ignore
all the others to a good approximation.

The driving rate of the $(n,k,m)=(3,1,2)$ r-mode for a polytrope of 
index 1 with
$M=1.4 M_{\odot}$ and $\rstar=12km$ is \citep{1999ApJ...521..764L}
\be
\gamma_{gr} & = & 0.05 \sec^{-1} \nu_{\rm khz}^6.
\label{eq:gammagr}
\ee

% -------------------------------------------------------------

\subsection{ bulk viscosity damping }

We now compute the damping rate of inertial modes by bulk viscosity
damping due to the modified URCA processes. We take the coefficient of
bulk viscosity from \citet{1989PhRvD..39.3804S} and 
\citet{1990ApJ...363..603C}.

The damping rate is
\be
- \dot{E}_{bulk} & = & \int d^3x\, \zeta \omega^2 |\grad \cdot \xivec|^2.
\ee
The Lagrangian compression is
\be
\grad \cdot \xivec & = & \frac{g\xi^r}{c^2} - \frac{\delta p}{\Gamma_1 p}
\simeq \frac{\xi^r}{H}
\ee
where the second equality is for low frequency modes.
The bulk viscosity coefficient is
\be
\zeta & = & \zeta_{fid} \omega^{-2} T_9^6 \rho^2
\ee
where
\be
\zeta_{fid} & = & (6 \times 10^{25}{\rm g\ cm^{-1} sec^{-3}} )
(10^{15} {\rm g\ cm^{-3}} )^{-2} 
\nonumber \\ & = &  6 \times 10^{-5} {\rm g^{-1} cm^5 sec^{-3}}.
\ee
Plugging in gives
\be
- \dot{E}_{bulk} & = & \zeta_{fid} T_9^6 
\int d^3x \,\frac{g^2 \rho^4 |\xi_r|^2}{\Gamma_1^2 p^2}.
\ee
We will evaluate this integral for a polytrope of index 1. In this case
\be
\frac{\Gamma_1 p}{\rho^2} & = & \frac{GM}{\rstar \rho_0}
\ee
is a constant so that 
\be
- \dot{E}_{bulk} & = & \zeta_{fid} T_9^6 \left(\frac{\rho_0 \rstar}
{GM} \right)^2
\int d^3x \,g^2 |\xi_r|^2.
\ee
In the WKB limit, this integral is logarithmically divergent at $r=\rstar$
and $\cos\theta=\pm |\mu|$.
This divergence implies that equal contributions to the 
integrand come per decade of distance from the surface. Since the true
eigenfunctions flatten off one wavelength from the surface, we cut off the
integrals at this distance.
Plugging everything into the integral
and approximating slowly varying quantities by their surface values
gives
the amplitude damping rate 
\be
\gamma_{bulk} & = & - \frac{\dot{E}_{bulk}}{2\Eunit}
= \frac{1}{8} \zeta_{fid} T_9^6 \frac{\rho_0}{r_*^2 \Omega^2} 
\frac{\ln \Lambda}{\mu^2},
\ee
where $\Lambda = 2|\mu|(1-\mu^2)^{1/2}p$ is roughly the number of nodes along
the rotation axis. Evaluating this expression for a fiducial neutron star
with polytrope index $n=1$, mass $M=1.4M_{\odot}$ and radius $\rstar=12 {\rm
km}$
we find the numerical value
\be
\gamma_{bulk} & = & 1.7 \times 10^{-10} \sec^{-1}\ T_9^6 \nu_{\rm khz}^{-2} 
\frac{\ln \Lambda}{\mu^2}.
\label{eq:bulk}
\ee
Note the extremely important fact that this damping rate is very weakly 
dependent on the wavelength of the mode! The usual picture of a cascade
to small scales does not make sense for damping by bulk viscosity. Instead
one must carry the energy to small frequency.

For the $(n,k,m)=(3,1,2)$ r-mode, the previously calculated value is
\citep{1999PhRvD..60f4006L}
\be 
\gamma_{bulk}(r-mode) & = & 2.8 \times 10^{-12} \sec^{-1} T_9^6
\nu_{\rm khz}^2.
\ee
The r-mode has a different scaling with $\Omega$ and normalization 
since the compression is smaller: $\grad \cdot \xivec \propto \Omega^2$
instead of $\grad \cdot \xivec \propto \xi^r/H$.

% -------------------------------------------------------------

\subsection{ shear viscosity }

The shear viscosity for nuclear matter has been calculated by 
\citet{1979ApJ...230..847F}, with an analytic fit by 
\citet{1990ApJ...363..603C} of the form
\be
\nu_s & = & \nu_{s,fid} (\rho/\rho_0)^{5/4} T_9^{-2}
\ee
where $\nu_{s,fid} = 2000\ {\rm cm^2 sec^{-1}} 
(\rho_0/10^{15}\ {\rm g\ cm^{-3}})^{5/4}$.

The shear viscosity damping is then
\be
-\dot{E}_{shear} & = & T_9^{-2} \int d^3 x \,\rho \nu_s \omega^2 \left( 
|\xi_{(i,j)}|^2 - \frac13 |\grad \cdot \xivec|^2 \right)
\nonumber \\ & \simeq & T_9^{-2}
\nu_{s,fid} \rho_0 \omega^2 \int d^3x (\rho/\rho_0)^{9/4} k^2 |\xivec|^2
\ee
where we have kept terms of leading order in $(|\kvec|H)^{-1}$,
and subscripted brackets denote a symmetrized derivative.
Plugging everything in, and approximating the density by a power law with 
depth appropriate for a polytrope of index 1, we find the damping rate
\be
\gamma_{shear} & \simeq  &
0.6 \pi \frac{p^2}{1-\mu^2} \frac{\nu_{s,fid}}{r_*^2} T_9^{-2}.
\ee
For our fiducial star this becomes
\be
\gamma_{shear} & = & 3.8 \times 10^{-9} \sec^{-1} T_9^{-2} \frac{p^2}{1-\mu^2}.
\label{eq:shear}
\ee

The previously computed r-mode shear damping rate is
\citep{1999ApJ...521..764L}
\be
\gamma_{shear}(r-mode) & = & 4 \times 10^{-9} \sec^ {-1} T_9^{-2},
\ee
which is about a factor of two different from our formula.

As first noted by \citet{2000ApJ...529L..33B}, the r-mode is damped much more
efficiently by shear in the crust-core boundary layer than by shear over 
the bulk of the stellar interior. \citet{2001MNRAS.324..917L}
later corrected this
damping rate to account for crust with a finite shear modulus. The key parameter
is the fractional velocity jump over the boundary layer, called $\eta$.
Levin and Ushomirsky found the rate of damping to be
\be
\gamma_{vbl}(r-mode) & = &
1.5 \times 10^{-3} \sec^{-1} \eta^2 \nu_{\rm khz}^{1/2} T_9^{-1} 
\ee
with a realistic estimate for the fractional velocity jump of
$\eta \sim 0.1$.
Inclusion of the finite shear modulus of the crust gives much
better agreement of the r-mode instability curve with the observations
of LMXB's. 

We have neglected damping of the daughter modes by shear in
the boundary layer.

% -------------------------------------------------------------

\section{ r-mode saturation by discrete modes: the small driving limit }
\label{sec:satd}

\begin{figure*}
\centerline{\psfig{figure=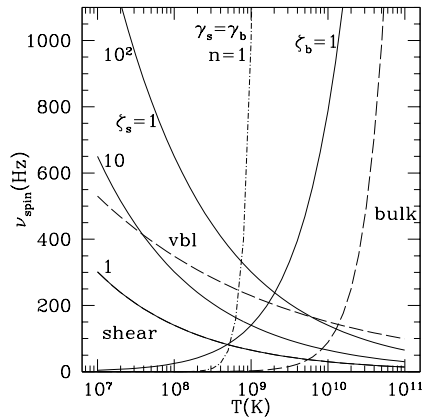,width=0.35\hsize}}
\caption[]{
Stability of energy transfer from the r-mode to daughter modes damped by
bulk or shear viscosity. The heavy dashed lines show where the r-mode is
marginally stable by equating $\gamma_{gr}$ to either $\gamma_{shear}$,
$\gamma_{vbl}$ for $\eta=0.1$,
or $\gamma_{bulk}$; the r-mode is
unstable above these lines.
The light solid lines show where $\gamma_{gr}$ equals the bulk ($\zeta_b=1$)
or shear ($\zeta_s=1$) viscosity damping of daughter modes.
The three lines $\zeta_s=1$ correspond to the number of nodes $n=1,10,100$
from bottom to top.
The stability of energy transfer to discrete daughter modes is only
stable in the region from the r-mode instability curve (dashed lines) to the
$\zeta_{b,s}=1$ lines, i.e., energy transfer to a small number of discrete modes
is not stable in the central portion of the r-mode instability region.
The nearly vertical dot-dashed line is where bulk viscosity damping equals
shear viscosity damping for large lengthscale ($n=1$) mode.
}
\label{fig:stability}
\end{figure*}

A fundamental plot for the r-mode instability is given in 
fig.\ref{fig:stability}. The r-mode is unstable for spin frequencies above
the thick dashed lines, where $\gamma_{gr}=\gamma_{shear}$, $\gamma_{vbl}$, or
$\gamma_{bulk}$. The solid lines show where driving of the r-mode equals
damping of daughter modes, indicating marginal stability of the energy 
transfer. For bulk viscosity, the ratio of driving to damping is
\be
\zeta_b & = & \frac{\gamma_{gr}}{\gamma_{bulk}}
= 8 \times 10^6 \nu_{\rm khz}^8 T_9^{-6}
\ee 
while for shear viscosity
\be
\zeta_s & = & \frac{\gamma_{gr}}{\gamma_{shear}}
= 1.3 \times 10^7 \nu_{\rm khz}^6 T_9^2 n^{-2}.
\ee
In these estimates we have used $\mu=1/6$, appropriate for daughter modes
with the largest coupling coefficients, and $n$ denotes the wavenumber
of the daughter mode.
Only in the region from the $\zeta_s=1$ and $\zeta_b=1$ lines to the r-mode
instability curve can we possibly have stable energy transfer for the three
mode system.

\subsection{ young neutron stars }

Nascent, rapidly rotating neutron stars cool into the region of
instability \citep{1998PhRvD..58h4020O} at fixed spin frequency.
For daughter modes mainly damped by bulk viscosity, there is a narrow region
near the instability curve in which energy transfer for a single triplet
of modes would be stable. 
However, in this region, the damping is relatively independent of
the daughter mode wavenumber. The quality factor of a daughter mode is roughly
\be
Q_d & \simeq & \frac{\omega}{\gamma_{bulk}}
\simeq 3.4 \times 10^5 \nu_{\rm khz}^3 T_{10}^{-6}
\ee
where we have used the daughter modes with the largest coupling coefficients
so that $\mu_{2,3}=1/6$. We have also set $\ln\Lambda \sim 1$. We found the 
coupling coefficients are roughly $\kappa \simeq 27$ for a parent mode
with $n_1=3$ and $\mu_1=1/3$ so that
the saturation amplitude for a three mode system is given by
\be
A_1^2(bulk) & = & \frac{E(r-mode)}{0.5 M \rstar^2 \Omega^2}
 \simeq  \frac{1}{\kappa^2 Q_d^2} 
\nonumber \\ & \simeq & 10^{-14} \nu_{\rm khz}^{-6} T_{10}^{12}.
\label{eq:dsatbulk}
\ee
This formula would imply that nascent neutron stars cooling
into the instability curve after a supernova will saturate at a
very small fraction of the rotational energy of the star.

However, this formula is not applicable for the following reason.
Since the damping rate of the daughter modes is relatively 
independent of the daughter mode wavenumber, all daughter modes have 
essentially the same parametric threshold (roughly eq.\ref{eq:dsatbulk})
until $n$ becomes large enough
that shear viscosity becomes comparable to bulk viscosity 
(see fig.\ref{fig:stability}). We estimate
this point to be at $n_{b=s} \simeq 10^4 T_{10}^4 \nu_{\rm khz}^{-1}$.
For daughter modes with frequency $\mu \sim 1/6$, {\it there are roughly 
$n_{b=s}^2 \simeq 10^8 T_{10}^8 \nu_{\rm khz}^{-2}$ daughter modes
  parametrically 
excited to large amplitude by
the parent r-mode, so that the discrete limit is not applicable.} Thus the
r-mode instability in young neutron stars is in the continuum limit
discussed in section \ref{sec:satc}.

\subsection{ LMXB's }

For the LMXB case, neutron stars with temperature $T \sim 3 \times 10^8K$ spin
up until they hit the instability curve \citep{1998ApJ...501L..89B,
1999ApJ...517..328L}. When the
instability curve near $T_9=0.3$ is set by boundary layer shear viscosity with
$\eta=0.1$ \citep{2001MNRAS.324..917L},
we see that if the star stays close to the instability curve,
one must go to daughter modes with $10-100$ nodes before
the energy transfer can become stable.
As a result, $10^{2-4}$ daughter modes will be
parametrically excited to large amplitude, and the continuum limit is
more appropriate.

If, however, boundary layer shear viscosity is not operating for some
reason, then the discrete mode approximation will be valid near the 
instability curve. The quality factor of the daughter modes in this case
is
\be
Q_d & \simeq & 5.5 \times 10^{9} \nu_{\rm khz} T_8^2 n^{-2}
\ee
giving a saturation amplitude for LMXB's near the instability curve to be
\be
A_1^2(shear) & = & \frac{E(r-mode)}{0.5 M \rstar^2 \Omega^2}
\simeq  \frac{1}{\kappa^2 Q_d^2}
\nonumber \\ & \simeq &  10^{-21} (0.33/\nu_{\rm khz})^2
 T_{8}^{-4}n^4,
\ee
which is quite small.

The conclusion we draw in this section is that, for the likely scenario
in which either bulk viscosity or boundary layer shear viscosity 
sets the r-mode instability curve, many modes will be
parametrically excited to large amplitude, and the continuum limit
discussed in the next section is a better approximation.

% -------------------------------------------------------------

\section{ r-mode saturation in the continuum limit}
\label{sec:satc}

The saturated r-mode energy was found in section \ref{sec:continuum} to be
\be
\frac{E(r-mode)}{0.5M\rstar^2\Omega^2}
& = &
10^{-6} \left( \frac{\alpha_e}{0.1} \right) \nu_{\rm khz}^5
\label{eq:saturationenergy2}
\ee
%
% EEF: comment about limitations of Fig 2 removed in next sentence
%
% Old version:
% This solution is valid when there is a clear separation between the 
% inner and outer scales of the turbulence (see fig.\ref{fig:cascade}, but notice
% that that drawing is only schematic, since the inner scale lines determined
% below involve both $\mu$ and $n$).
%
This solution is valid when there is a clear separation between the 
inner and outer scales of the turbulence (see Fig.\ \ref{fig:cascade}).
The outer scale is given by the r-mode itself, while the inner scale 
is where $\gamma_{\rm nl}$, the characteristic rate for amplitude change
by nonlinear interactions, is equal to the dissipation rate given by
$\gamma_{\rm shear}$ (bulk viscosity is irrelevant for the inner
scale; see below).  In other words, the 
inner scale is where the Reynolds number for that scale becomes order unity.
We can
estimate the nonlinear interaction rate using eqs.\ref{eq:ndot} 
and \ref{eq:kinetic}, with the scalings $n = n_\alpha \sim n_\beta \sim
n_\gamma$, $ \mu = \mu_\alpha \sim \mu_\beta \sim \mu_\gamma$, etc.
We find
\be
\gamma_{\rm nl} & \sim & \frac{I}{N} 
\simeq \frac{1}{N} \frac{n^5}{\mu} (\Omega N)^2 
\sim \gamma_{\rm gr} \frac{n}{n_r} 
\left(\frac{\mu_r}{\mu} \right)^{3/2}
\label{eq:gammanl}
\ee
where $n_r$ and $\mu_r$ set the scale for the driving region at 
which $\gamma_{\rm gr}=\gamma_{\rm nl}$.

%
%  EEF: I've modified the discussion in the following two paragraphs,
%  and also interchanged their order.
%

The expression for shear viscosity from eq.\ref{eq:shear} can be
written $\gamma_{\rm shear}= \gamma_s n^2$ in the $\mu^2 \ll 1$
limit. Equating $\gamma_{\rm nl}$ to $\gamma_{\rm shear}$, we find the 
inner scale is 
\be
 \frac{n}{n_r} \left( \frac{\mu}{\mu_r} \right)^{3/2} 
& \simeq  &
n_r^{-2} \frac{\gamma_{\rm gr}}{\gamma_s}
\nonumber \\ & \simeq & 
170\ n_r^{-2}\,
\left( \frac{\nu_{\rm spin}}{330\ {\rm Hz}} \right)^6 T_8^2.
\label{eq:shearregime}
\ee
In the region of the $(\Omega, T)$ plane where the right hand side of
Eq.\ (\ref{eq:shearregime}) is large, many modes are excited with a
clear separation between inner and outer scales of turbulence (see
Fig. \ref{fig:cascade}).  This region includes the entire instability
window when the r-mode is damped by a viscous boundary layer.   When
there is no viscous boundary layer, there is a small region close to
the instability curve where (\ref{eq:shearregime}) is small and 
where the discrete limit applies instead of the continuum limit.

For bulk viscosity things work differently.  
The cascade solution will be valid in the region of phase space where
$\gamma_{\rm nl} \simgreat \gamma_{\rm bulk}$, or
\be
\frac{\mu}{\mu_r} \left(\frac{n}{n_r} \right)^2  & \simgreat & 
\mu_r^{-4} \left( \frac{\gamma_b}{\gamma_{\rm gr}} \right)^2
\nonumber \\ & =  & 
10^{-5}\ \mu_r^{-4}\, T_{10}^{12}\, \nu_{\rm khz}^{-16},
\label{eq:bulkregime}
\ee
where we have written $\gamma_{\rm bulk}=\gamma_b \mu^{-2}$.
In the region of the $(\Omega,T)$ plane where the right hand side of
Eq.\ (\ref{eq:bulkregime}) is large compared to unity, bulk viscosity
is dominant at the outer scale and no cascade solution exists.
When the right hand side of Eq.\ (\ref{eq:bulkregime}) is small
compared to unity, then a cascade can form, but the bulk viscosity is
irrelevant for setting the inner scale of the cascade.  We note that
the boundary (\ref{eq:bulkregime}) of the bulk viscosity
dominated regime approximately coincides with the curve $\zeta_b=1$ in
Fig.\ \ref{fig:stability}.  A newly born neutron star will very
rapidly move from the instability curve to the $\zeta_b=1$ curve at
which point the cascade can form.

Finally we note 
\footnote{ We thank P. Goldreich for bringing this to our attention.}
that the weak turbulence approximation which underlies
the derivation of Eq.\ (\ref{eq:ndot}) eventually breaks down as one goes
to small scales.  The weak turbulence approximation requires that the
nonlinear energy transfer timescale $1/\gamma_{\rm nl} \sim 20 \, {\rm
s} \, \nu_{\rm khz}^{-6} n^{-1} \mu^{3/2}$ be much longer than the mode
period $\sim 5 \times 10^{-4} \, {\rm s} \, \mu^{-1} \nu_{\rm khz}^{-1}$,
which breaks down in the regime \begin{equation} n \mu^{-5/2} \simgreat
4 \times 10^4 \, \nu_{\rm khz}^{-5}.  \label{weakturbulencefailure}
\end{equation} Thus our turbulent cascade solution of the equations of
motion (\ref{eq:ndot}) will likely be replaced by some form of strong
turbulence at sufficiently small scales.  However, this should not affect
our prediction of the r-mode's saturation amplitude, as the regime
(\ref{weakturbulencefailure}) in phase space where the approximation
breaks down is well separated from the driving regime $\mu \sim n \sim 1$.

% -------------------------------------------------------------

\section{comparison with previous work}
\label{sec:comparison}

There have been three distinct alternative nonlinear mechanisms proposed 
to saturate the growth of the r-mode:
(1) For large amplitude pulsations, the Fermi energies of the electron-
proton-neutron gas become significantly shifted, and the kinetic energy
can be rapidly converted to both heat and neutrinos by nonlinear bulk
viscosity \citep{reisenegger}.
(2) The amplitude grows so large
%
%  EEF: corrected the following formula for gravitational binding
%  energy of the star
%
($E \sim E_{\rm rotation} \sim GM^2/\rstar$) 
that strong shocks occur, rapidly thermalizing the kinetic energy
\citep{2001PhRvL..86.1152L,lindblomv2};
(3) In neutron stars with
a crust, a turbulent boundary layer forms at the crust-core interface
\citep{2001ApJ...549.1011W}.
We now discuss each in a bit more detail.

Since the Fermi energy of the electron and neutron have a different 
dependence on density, the Fermi surfaces are shifted out of beta equilibrium
when matter is compressed. The scaling of the resulting neutrino emission
rate depends on the ratio of chemical potential imbalance,
$ \Delta \bar{\mu}_\nu=\mu_n - \mu_p - \mu_e $, to temperature, which
is roughly \citep{1995ApJ...442..749R}
\be
\frac{ \Delta \bar{\mu}_\nu }{T} & \simeq  & \frac{1}{3} \frac{E_{F,e}}{T}
\frac{\Delta \rho}{\rho}
\ee
where $E_{F,e}$ is the Fermi energy of the electron. When this ratio
is large, $5/8$ of the resulting dissipation
heats the star and $3/8$ goes into neutrinos. The rate of such dissipation
scales as $\left( \frac{ \Delta \bar{\mu}_\nu/T}{7.33}  \right)^8
\propto (\Delta \rho/\rho)^8$ times
the neutrino emissivity of uncompressed matter. The mode damping rate is
extremely sensitive to the compression, and can saturate the growth
of the r-mode for sufficiently large amplitude. \citet{reisenegger}
has done a detailed calculation, finding that
the saturation energy is comparable to the stellar rotation energy.
This interesting idea gives a larger (less constraining) saturation 
amplitude compared to the
value found in this paper.

Next, \cite{ 2001PhRvL..86.1152L,lindblomv2} 
have performed state-of-the-art 3D Newtonian 
hydrodynamics simulations including a prescription for the radiation
reaction force. The only damping mechanism included in the code is 
numerical viscosity, and of order $128^3$ points were used. They were able
to follow the linear growth of the r-mode, all the way into the nonlinear
regime where $E \geq E_{rot}$. Shocks then formed near the surface of the
star, rapidly thermalizing the kinetic energy of the mode.

Since the
growth of the instability is so slow compared to the dynamical time in the
star, they found it necessary to artificially
increase the radiation reaction force
by a factor of $\sim 4500$.  
A natural question is how the mode would saturate if the correct, physical
value of the driving force was used. The following physical example is
useful to consider.
Imagine water waves being driven by wind moving at 
$1\ {\rm cm\, sec^{-1}}$, a whisper of a breeze, as compared to 
$4500\ {\rm cm\, sec^{-1}}$, a hurricane. For small amplitude water waves, four
wave interactions can transport energy to small scales, saturating
the growth of the waves. In a hurricane,
the wave growth time is so short that waves grow to large amplitudes and
break. Since Lindblom et al. have
not addressed how saturation might occur for physical values of either
driving or damping of the waves, the relevance of their simulations
to saturation of the r-mode instability is not clear.

One comparison which can be made is to use our formula in 
eq.\ref{eq:saturationenergy1} to estimate the saturation amplitude seen
in Lindblom et al.'s simulations when $\gamma_{gr}
\rightarrow 4500 \gamma_{gr}$. We find
$E_{\rm simulation}/(0.5M\rstar^2\Omega^2)
 = 0.1 \left(\alpha_e/0.1 \right) (4500 \gamma_{gr}/\Omega)
\simeq 5 \times 10^{-3} \left(\alpha_e/0.1 \right) \nu_{\rm khz}^5$.
This result can be translated into Lindblom et al.'s notation by using
$A_1^2 \simeq 0.5 {\tilde J} \alpha_{\rm simulation}^2 = 0.008 \alpha_{\rm
simulation}^2$ with the result  
$\alpha_{\rm simulation} \simeq 0.7 \left(\alpha_e/0.1 \right)^{1/2}
\nu_{\rm khz}^{5/2}$. This comparison shows that if one attempted to 
extrapolate the saturation amplitude over more than three decades in 
driving force, that the saturation amplitude by mode coupling would
indeed be of order unity, in their notation. This does not, of course, explain
the saturation amplitude seen in Lindblom et al.'s simulations, which they
explain is due to strong shocks near the stellar surface.
However, the results of this paper show that their claims of (1) saturation
energy of order the rotation energy, and (2) strong shocks as the saturation
mechanism, are not supported. 
They are an artifact of the unphysically large value
for the radiation reaction force.

We comment further on
the ability of a numerical simulation to accurately reproduce the cascade
of energy to small scales as derived in this paper. In simulations with
$128^3$ points, only a certain number of modes exist because of the finite
resolution. Since the detuning is a rapidly decreasing function of wavenumber,
secular energy transfer by nearly resonant interactions becomes
more important as the number of grid points increases. For instance,
daughter modes with half the frequency of the r-mode have quantum numbers
in the ratio $k/n \simeq 1/20$, so that one needs nearly $20$ times as
many nodes in the cylindrical radius as in the z direction in order to
find parametrically excited daughter modes. We estimate that only a few of
these might have been accurately modeled by Lindblom et al's simulations.
In time evolutions of the mode amplitude equations with a small number of low
order, nonresonant modes \citep{sharontalk,philyammer}, large saturation 
amplitudes were found as compared to the results in this paper. The reason,
as can be clearly seen in eq.\ref{eq:parametric} for the parametric  
threshold, is that the r-mode cannot easily excite daughter pairs with large
detuning. However, going to higher order modes with much smaller detuning
can give a saturation amplitude orders of magnitude smaller than for 
arbitrary, low order modes.

Lindblom et al. specifically commented that three-mode coupling is not
the saturation mechanism in section H of their paper. Their claim was based
on the lack of power observed in certain modes besides the r-mode during their
simulation. However, they focused on interactions which couple the r-mode
twice to a third mode. As they themselves comment at the end of section H,
they have not included parametric excitation of daughter modes in their
constraints. As discussed in section \ref{sec:discrete}, couplings of the 
type discussed by \citet{lindblomv2} are far less important than parametric
couplings, because (1) they are down by a factor of parent mode amplitude, 
which is small, and (2) only a relatively small region of phase space 
couples well with the r-mode by non-parametric couplings.
Hence, Lindblom et al.'s constraints are not
useful since they constrain an unimportant process. 

%
% EEF: added the following paragraph following John Friedmans'
% email.
%

The inability of simulations to include very high order modes
presumably also explains the results of the fully relativistic
simulations of \cite{FS2001}, in which an r-mode with order unity
amplitude was observed not to lose any energy to other modes over
several dynamical times.  More recent simulations by
\cite{Gressman2002} show that for slightly larger initial amplitudes, the
r-mode decays rapidly into a differentially rotating configuration
without shocks forming.  These results are not inconsistent with our
analyses, but our results indicate that the r-mode never reaches the
regime of rapid nonlinear decay seen by \cite{Gressman2002}.

Next we discuss the turbulent boundary layer mechanism of
\cite{2001ApJ...549.1011W} which operates in neutron stars with a crust.
Energy dissipation by turbulent drag 
scales as $A_1^3$, leading to saturation of the mode.
The attractiveness of this idea is that the turbulent drag force is well
understood in magnitude and scaling both from numerical estimates as well
as laboratory experiments. These authors considered the effect of such 
energy dissipation on the crust and thermal history of the star, and
go on to discuss the observable spin frequency
of the star after it exits the r-mode instability region.
% Before this paper,
%their paper represents the only attempt to discuss mode
%saturation by including relevant physical effects such as damping mechanisms.
For a realistic fractional velocity jump
across the crust-core boundary layer, $\eta \sim 0.1$, they found the
r-mode saturated at a value $E/(0.5M\rstar^2\Omega^2) \simeq 0.2
\nu_{\rm khz}^{10}$,
which is larger (less constraining) than the value found here, both
in normalization, and in the dependence on $\nu_{\rm khz}$. Furthermore, their
mechanism does not operate in completely fluid stars without a crust,
which is the case for hot young neutron stars.

Lastly, we mention that this paper is a companion paper to that of 
\citet{morsink2002}, which discusses the nonlinear coupling among r-modes
in a star for which buoyancy forces are dominant over Coriolis forces.
Morsink found that, because the r-mode frequency decreases with $m$, 
interactions do not become more resonant as the daughter mode $m$ increase.
As a result, energy transfer among three r-modes is not likely to produce
a saturation value as low as in this paper.

We conclude that nonlinear mode coupling to inertial modes
provides the most stringent 
constraints on the r-mode amplitude at this time.

% -------------------------------------------------------------

\section{ spin evolution of neutron stars}
\label{sec:evolution}

The spindown torque exerted on the neutron star by gravitational radiation is
roughly 
\be
\tau_{gr} & \simeq & 0.12 \left( \frac{\Omega \rstar}{c} \right)^7
\frac{GM^2}{\rstar} A_1^2
\nonumber \\ & \simeq & 
 2 \times 10^{42}\, {\rm erg}\, \left( \frac{\alpha_e}{0.1} \right)
 \nu_{\rm khz}^{12}.
\ee
The spindown time associated with this torque is
\be
t_{\rm spindown} & \simeq & \frac{M \rstar^2 \Omega}{11 \tau_{gr}}
\simeq 0.01\,{\rm yr}\, \left( \frac{\alpha_e}{0.1} \right)^{-1}
\nu_{\rm khz}^{-11}
\nonumber \\ & \simeq & 
 2 \times 10^3\, {\rm yr}\, \left( \frac{\alpha_e}{0.1} \right)^{-1} 
\left( \frac{330\,\rm Hz}{\nu} \right)^{11}.
\label{spindowntime}
\ee
%We can conclude that the r-mode instability is able to spin rapidly
%rotating neutron stars down to the point that they exit the instability 
%curve within of order $10^3 $ years.
Since the spindown rate decreases
strongly with spin frequency, most of the time is spent at lower 
frequencies.

The spindown time becomes $\simgreat 10^4\ {\rm yr}$ at the lowest
rotation rates inside the instability curve, while it is of order a few 
days for stars rotating near breakup. This is of interest for certain
gamma-ray burst models, such as the ``supranova" model
\citep{1999ApJ...527L..43V} in which core collapse leads to ejection
of the stellar envelope, as well as a rapidly rotating neutron star
which is {\it above} the maximum mass for a nonrotating star. Angular momentum
transport can then slow the neutron star down, leading to collapse to a black
hole and generation of a powerful gamma-ray burst.
Our results imply that the gamma-ray
burst should occur within of order a week after the supernova explosion.

Next we turn our attention to neutron stars in LMXB's. 
The ratio of spindown torque, due to radiation reaction, to accretion torque
$\tau_{acc} \simeq \dot{M} (GM\rstar)^{1/2}$ is roughly
\be
\frac{\tau_{gr}}{\tau_{acc}} & \simeq & \frac{ 2 \times 10^{42} 
(\alpha_e/0.1) \nu_{\rm khz}^{12}\ }
{ 10^{34} (\dot{M}/ 10^{-8}\ M_{\odot}\ {\rm yr^{-1}}) }
\nonumber \\ & \simeq &
\frac{ 10^{-8} M_{\odot}\ {\rm yr^{-1}} }{\dot{M}}
\left( \frac{\alpha_e}{0.1} \right)
\left( \frac{\nu_{\rm spin}}{200 \,\rm Hz} \right)^{12}.
\ee
For accretion rates smaller than the Eddington rate $\dot{M} \simeq
10^{-8}\ M_{\odot}\ {\rm yr}^{-1}$, and spin frequencies above
$\nu_{\rm spin}=200\,{\rm Hz}$, the radiation reaction torque is larger than
the accretion torque and can halt the further spinup of the neutron star.
%
% EEF: removed foonote in the following sentence since the equilibrium
% LMXB case now discussed in the next section
%
If the neutron star viscosity is dominated by normal matter, then the
star enters into a limit cycle of spinup by accretion 
and spindown by the r-mode, as discussed by
\citet{1999ApJ...517..328L}.  [The alternative equilibrium scenario is
discussed in Sec.\ \ref{sec:detection} below.]  Since the r-mode is only likely
to be unstable for $\nu_{\rm spin} \simgreat 
300 \,{\rm Hz}$, the r-mode can halt spinup inside the region of instability.

The observable spin frequency is determined by where the star exits the 
region of r-mode instability, if no other process spins the star down
further.
The exact spin frequency at which the star exits the region of r-mode 
instability depends on the evolution of both the spin frequency and the
stellar temperature \citep{1999ApJ...517..328L,1998PhRvD..58h4020O,
2001ApJ...549.1011W}. We can estimate this terminal frequency
\citep{2001ApJ...549.1011W} by equating the neutrino cooling luminosity,
$L_\nu = 7.4 \times 10^{39}\ T_9^8\, {\rm erg\, sec}^{-1}$,
with the rate of stellar heating due to the r-mode.
If we approximate that all the energy input to
the r-mode by radiation reaction is damped away as heat, the rate of 
heating of the star is just given by $\dot{E}_{\rm heat} = 2 \gamma_{gr}
E$, where $E$ is the saturation energy found in eq.\ref{eq:saturationenergy2}.
Equating heating and cooling, we find the equilibrium temperature as a function
of spin frequency, given by
\be
T_{\rm eq} & \simeq & 10^9\ K\ \left( \frac{\alpha_e}{0.1} \right)^{1/8}
\left( \frac{\nu_{\rm spin}}{330 \,\rm Hz} \right)^{13/8}.
\ee
The crystallization temperature of the crust is \citep{2001ApJ...549.1011W}
$ T_{\rm melt}
\sim (5-10) \times 10^9\, K$
so the heating by the r-mode cannot prevent the crust from forming when 
$\nu_{\rm spin} \ll 10^3\, {\rm Hz}$.
If the instability curve is set by boundary layer shear viscosity ($\gamma_{gr}
= \gamma_{vbl}$), the intersection of the equilibrium spin down curve with
the r-mode instability curve is given by the terminal frequency
\be
\nu_{\rm terminal} & \simeq & 250\ {\rm Hz}\ 
\left( \frac{\alpha_e}{0.1} \right)^{0.02} 
\left( \frac{\eta}{0.1} \right)^{0.28}
\label{eq:terminalnu}
\ee
with a core temperature of roughly 
\be
T_{\rm terminal} & \simeq & 6 \times 10^8\ K\ 
\left( \frac{\alpha_e}{0.1} \right)^{0.15}
\left( \frac{\nu_{\rm spin}}{330 \,\rm Hz} \right)^{0.46}.
\ee
Note that the observable spin frequency is very insensitive to the
saturation parameter $\alpha_e$, as well as to the fractional velocity 
jump $\eta$. The spin frequency found in eq.\ref{eq:terminalnu} is 
comparable to the lower end of the observed LMXB's, consistent with 
 a limit cycle \citep{1999ApJ...517..328L} of spin-up
by accretion and spin-down by the r-mode.
The timescale to exit the instability curve is roughly $t_{\rm spindown}
\simeq 2000\ {\rm yr}\ (330/250)^{11}\ (\alpha_e/0.1)^{-1} \simeq 
4 \times 10^{4}\ {\rm yr}\ (\alpha_e/0.1)^{-1}$. This spindown timescale
is very sensitive to the position of the instability curve.

For young neutron stars with strong magnetic fields, the spindown torque
from magnetic dipole radiation is comparable to that from gravitational 
radiation. Equating the magnetic dipole spindown timescale \citep{1983bhwd.book.....S}
$t_{\rm md} \simeq 30\, {\rm yr}\, B_{12}^{-2} \nu_{\rm khz}^{-2}$ to $t_{\rm 
spindown}$, we find that gravitational radiation reaction dominates
for frequencies above $\nu_{\rm spin} \simeq 400\, {\rm Hz}\, (0.1/\alpha_e)^{1/9}\,
B_{12}^{2/9}$, where $B_{12}$ is the surface dipole field in units of $10^{12}\ 
{\rm G}$. Hence for typical pulsars with magnetic fields $\sim 10^{12}
{\rm G}$, the spindown torque is dominated by the r-mode only for fairly large
spin frequencies.

\section{Detectability of gravitational waves}
\label{sec:detection}

We now discuss the prospect of detecting gravitational waves from
r-modes, based on the saturation amplitude (\ref{eq:continuumsat}).
We consider three different scenarios: (i) newly born neutron stars
where an optically observed extra-Galactic supernova provides the sky
location for the gravitational wave search; (ii) LMXB's in the
spinup-spindown limit cycle first discussed by
\citet{1999ApJ...517..328L}; and (iii) LMXB's in spin and thermal equilibrium.

For newly born neutron stars, \citet{2000PhRvD..612001B} (BC) discuss the
detection likelihood by LIGO assuming a large saturation amplitude.   
They parameterize the saturation amplitude in terms of a parameter
$\kappa$ in their Eq.\ (7.3), which we denote by $\kappa_{bc}$.  Our
result (\ref{eq:continuumsat}) gives $\kappa_{bc} \simeq 1.2 \times
10^{-3} \nu_{\rm khz}^5$, while BC took $\kappa_{bc}=1$.  In the first year
of spindown, $\nu_{\rm 
khz}$ decreases from $\sim 1$ to $\sim 0.66$ [cf.\ Eq.\
(\ref{spindowntime}) above], and thus the gravitational wave strain
amplitude will be a factor $\sqrt{\kappa_{bc}} \sim 2.8 \times 10^{-2}$
smaller than that considered by BC.  The distance to which the source
can be seen by enhanced LIGO detectors, for fixed integration time
(see below), is
correspondingly reduced from 
BC's estimate of $\sim 8 \, {\rm Mpc}$ to $\sim 200 \, {\rm kpc}$, 
almost inside the Galaxy. Since the 
galactic supernova rate is roughly once per $50-100\ {\rm yrs}$, the
probability that LIGO will detect young neutron stars radiating due to
r-modes is small.

We now discuss why we can treat the integration time as fixed.
The matched filtering signal to noise ratio $S/N$ for gravitational waves,
when averaged over source orientations and polarizations, depends only
on the energy per unit frequency $dE/df$ of the waves \citep{FH98}:
\begin{equation}
{S^2 \over N^2} = {2 G \over 5 \pi^2 c^3 D^2} \,
\int df \, {1 \over f^2 S_h(f)} \, {d E \over d f}.
\label{snraveraged}
\end{equation}
Here $D$ is the distance to the source and $S_h(f)$ is the detector
noise spectrum.  For waves of fixed azimuthal quantum number $m$,
using the replacement $dE = 2 \pi f dJ / m$ yields 
\citep{blandford84,lindblomowen2002b}
\begin{equation}
{S^2 \over N^2} = {4 G \over 5 \pi m c^3 D^2} \,
\int df \, {1 \over f S_h(f)} \, {d J \over d f},
\label{snraveraged1}
\end{equation}
where $J$ is the $z$-component of angular momentum.
As noted by \cite{lindblomowen2002b}, the expression (\ref{snraveraged1}) is
independent of how quickly the star looses angular momentum, and hence
of the saturation amplitude.  Thus, {\it a priori} one would not expect our low
saturation amplitude (\ref{eq:continuumsat}) to affect very much the
detectability of the signal.  The problem however is that it is
not possible to integrate long enough to accumulate the total
signal-to-noise ratio (\ref{snraveraged1}).

Using the stellar model discussed before Eq.\ (\ref{eq:gammagr}), the
relation $f = 4 \nu_{\rm spin}/3$ between gravitational wave frequency $f$
and spin frequency, the broadband LIGO-II noise curve \footnote{In
the relevant frequency range $f \simgreat 500 \, {\rm Hz}$, this noise curve
is approximately given by $f S_h(f) \simeq 1.7 \times 10^{-44} (f / 1000
\, {\rm Hz})^{3}$.} from \cite{LIGOII}, and neglecting the spin
dependence of the moment of inertial of the star,
we can evaluate
(\ref{snraveraged1}) for a spindown from an 
initial spin frequency $\nu_{\rm khz,i}$ in kHz to a final spin frequency
$\nu_{\rm khz,f}$.  The result is
\begin{equation}
{S \over N} = 5.4 \left( {10 {\rm Mpc} \over D} \right) \left(
{ 1 \over \nu_{\rm khz,f}^2} -{ 1 \over \nu_{\rm khz,i}^2} \right)^{1/2}.
\label{snrans}
\end{equation}
The complete spindown from say $1 \, {\rm kHz}$ to $\sim 250 {\rm Hz}$
[cf.\ Eq.\ (\ref{eq:terminalnu}) above] gives $S/N \sim 21$ at $10 \,
{\rm Mpc}$.
The first year of spindown from $1000 \, {\rm Hz}$ to $650 \,
{\rm Hz}$ [cf.\ Eq.\ (\ref{spindowntime}) above] gives instead $S/N
\sim 6.2$, which is not too much smaller.

However, the need to perform a search over spindown parameters in
practice limits the integration time to $\sim 10^6$ seconds.
BC analyzed the performance of a ``stack
slide'' search method, involving demodulating the signal for many
different choices of spindown parameters, dividing the demodulated
data into several chunks or ``stacks'', computing the power spectrum
of each stack and adding the power spectra.  
The threshold value $\rho_{\rm th}$
of $S/N$ for this method, assuming $1\%$ false alarm probability, is
approximately given by 
solving the equation 
\begin{equation}
\Gamma(N_s,N_s + \rho_{\rm th}^2/4) / \Gamma(N_s) = 0.01 / (N_b N_p).
\label{threshold}
\end{equation}
where $\Gamma(a,b)$ is the incomplete gamma function, $N_s$ is the
number of stacks, $N_b$ is the number of frequency bins per stack, and
$N_p$ is the number of points in the space of spindown parameters.
BC's estimate that r-modes are detectable out to $8 \, {\rm Mpc}$ 
was based on assuming a Teraflop of computing
power, which implied an optimum detection strategy
of $\sim 8$ stacks of $\sim 10^5$ s duration each,
integrating from $\nu_{\rm spin} = 200 \, {\rm Hz}$ to $\nu_{\rm spin}
\sim 186 \, {\rm Hz}$.

We can modify the BC analysis for our turbulent cascade scenario as
follows.  Optimum sensitivity is achieved late in the spindown, so we assume
that $\nu_{\rm spin} \sim 650 \, {\rm Hz}$, corresponding to 1 year
after the start of the spindown if the initial spin frequency is 1 kHz.
We take the parameter values $\mu_{\rm max} = 0.3$, $f_{\rm max} =
(4/3) 650 {\rm 
Hz}$ (instead of $200 \, {\rm Hz}$ as in BC), and a spindown 
timescale $\tau_{\rm min} = 1 \, {\rm yr}$, which from Eq.\
(\ref{spindowntime}) is appropriate after 1 year of spindown.
Maximizing over the number of stacks and stack durations as in BC
gives that the optimum detection strategy for a Teraflop of computing
power is to use $\sim 10$ stacks of duration $\sim 3 \times 10^4$ s
each.  The corresponding number of parameter space points is $N_p \sim
6 \times 10^8$, from Eq. (2.20) of BC, which gives from Eq.\ (\ref{threshold})
a threshold value of $\rho_{\rm th} \sim 15$.  Combining this with
Eq.\ (\ref{snrans}), and noting that at $\nu_{\rm spin} \sim 650 {\rm
Hz}$ an integration time of $3 \times 10^5$ s corresponds to $\Delta \nu_{\rm
spin} \sim 0.5 (\alpha_e / 0.1) \, {\rm Hz}$ gives that the source
would be detectable to 
$\sim 200 (\alpha_e / 0.1)^{1/2} \, {\rm kpc}$, consistent with our
earlier estimate\footnote{  
The main reason for the loss of sensitivity compared to BC is the reduction in
$\Delta \nu_{\rm spin}$ from $\sim 14 \, {\rm Hz}$ to $0.5 \, {\rm
Hz}$; the star is spinning down more slowly.}.

This conclusion, however, is based on the assumption of using the
stack-slide search method.  It is conceivable that an alternative 
signal processing strategy (and increased computational power) might
enable one to integrate for longer periods and achieve a sensitivity
closer to the original BC estimate.

We mention in passing another possible difficulty in searching for the signal
from r-modes when a turbulent cascade is present.  
This difficulty is that the phase of the r-mode will wander randomly
in time due to the interaction with the turbulent cascade, on some
timescale $t_{\rm c}$.  The peak in the Fourier transform of the
demodulated data stream will correspondingly be smeared out over a
frequency interval of width $\sim 1/t_{\rm c}$, which will be over
several frequency bins if the stack size is larger than $t_{\rm c}$.

The phase coherence timescale for a typical mode in
the cascade will be of order 
\beq
t_c \sim {1 \over \gamma_{\rm nl}},
\label{ctimescale}
\end{equation}
or smaller, where $\gamma_{\rm nl}$ is the the nonlinear energy
transfer rate (\ref{eq:gammanl}) \citep{1992kstw.book.....Z}.  For the
r-mode this is only $\sim 200 \, {\rm s}$ at $\nu_{\rm spin} \sim 700
{\rm Hz}$.  However, one might expect the coherence time of the r-mode
to be somewhat longer than the estimate (\ref{ctimescale}), since the
r-mode is being pumped coherently and is loosing energy by interacting
simultaneously with a large number of different modes.  
Unless the phase coherence time for the r-mode is $10^2 - 10^3$ times larger
than the estimate (\ref{ctimescale}), the sensitivity of the search
will be reduced.  Again, it may be possible to modify the data
analysis procedure to compensate for the phase wandering 
%(for
%example, by using as a statistic the total power in several adjacent
%frequency bins)
\footnote{The method suggested by BC to compensate for
phase wandering requires a stack size shorter than $t_c$ and a
computational power that scales as $3^{T/t_c}$, where $T$ is the total
integration time.  In practice this limits $T$ to $\simless \, (20 -
30) t_c$.}.

Next, we consider the detectability of r-modes in LMXB's in the
spin up/spin down limit cycle.  The signal from the spin down phase is
essentially the same as for newborn neutron stars, except that they
will typically be seen at a low frequency where most of the spindown
time is spent.  At lower spin frequencies the search over spin-down
parameters becomes significantly easier, since the spindown timescale
is longer.  The formula (4.3) in BC for computational power,
with $f_{\rm max} = \nu_{\rm spin}$ and $\tau_{\rm min}$ given by Eq.\
(\ref{spindowntime}), shows that for $\nu_{\rm spin} \simless 400 \,
{\rm Hz}$ integration times as long as $10^7$ s can be achieved with
1 Teraflop of computing power.  Combining Eqs.\ (\ref{spindowntime})
and (\ref{snrans}) gives that the signal to noise ratio for a $10^7$ s
integration starting at $\nu_{\rm spin}$ is
\begin{equation}
{S \over N} \sim 9 \left( { \nu_{\rm spin} \over 330 \, {\rm Hz} }
\right)^{9/2} \, \left( {100 \, {\rm kpc} \over D} \right) \, \left( {
  \alpha_e \over 0.1} \right)^{1/2}.
\end{equation}
In the regime $\nu_{\rm spin} \lesssim 400 {\rm Hz}$, the signal-to-noise
threshold from the BC method is $\rho_{\rm th} \sim 10$, within a
factor of $\sim 2$, giving that the signal should be visible to a
distance
\begin{equation}
D \sim 90 \, {\rm kpc} \, \left( { \nu_{\rm spin} \over 330 \, {\rm Hz} }
\right)^{9/2}  \left( {  \alpha_e \over 0.1} \right)^{1/2}
\label{dist1}
\end{equation}
for $\nu_{\rm spin} \lesssim 400 \, {\rm Hz}$.

However, as noted by \citet{1999ApJ...517..328L}, the chance of observing a
particular source emitting gravitational waves is proportional to the
relative length of time spent in the spindown phase of the limit cycle. 
Using our saturation amplitude we find a duty cycle $\sim 10^{-3} (0.1
/ \alpha_e)$, implying that one would need of order $10^3 (\alpha_e /
0.1)$ LMXB's within the distance (\ref{dist1}) in order to overcome the small
duty cycle.  Nevertheless, as argued by \cite{Heyl2002}, there may be
enough Galactic LMXB's that some will be seen in the spin-down phase
by enhanced LIGO, especially if $\alpha_e$ is smaller than $0.1$.

We note that for LMXB's, the phase wandering of the r-mode due to
the turbulent cascade is less of a problem, since the nonlinear energy transfer
timescale (\ref{eq:gammanl}) increases rapidly as $\nu_{\rm spin}$
decreases.  There is in addition a phase wandering due to fluctuations
in the accretion torque, but this occurs over much longer timescales 
and can be dealt with in the manner suggested by BC.

The third possibility we consider is when 
the viscosity of an accreting star is independent of temperature or is an
increasing function of temperature.  In such a case the star can
achieve an equilibrium state where the accretion 
spin-up torque is stably balanced by the radiation reaction torque due
to the r-mode, and r-mode heating is balanced by neutrino cooling
\citep{1999ApJ...517..328L}.    
Such equilibria have been found for stars with hyperon cores 
\citep{Wagoner2002} and for strange stars \citep{AJK}.
[However, the central densities of neutron stars are sufficiently uncertain
that hyperon cores may or may not exist.]
In these scenarios, the equilibrium r-mode amplitude
is not set by the turbulent cascade considered here, but instead by
the equilibrium conditions.  The gravitational wave signal is weaker
than the limit cycle case considered above, and its strength can be
inferred from 
the X-ray flux \citep{1998ApJ...501L..89B}.

\section{ conclusions }
\label{sec:conclusions}

In this paper, we have accomplished several objectives, which can be
divided into stellar oscillation theory, and phenomenology of neutron
star spin evolution.

We have, for the first time, presented a WKB theory of global stellar inertial 
modes (section \ref{sec:whichmodes} and appendix \ref{app:kappa}),
including both the rapidly varying phase and amplitude which rises
quickly toward the surface. Both the eigenmodes and pulsation frequencies
take on a very simple form, which was never clearly elucidated in 
previous calculations [e.g., \citet{1999PhRvD..59d4009L,1999ApJ...521..764L}.]
Appendix \ref{app:kappa} gives a detailed mathematical treatment of inertial
waves.
We have estimated when the affects of
buoyancy become important. The damping rates by bulk and shear viscosity
appropriate for neutron stars have been derived in section \ref{sec:damping},
and reduced to simple, accurate formulae giving the scalings with
stellar parameters and mode quantum numbers.

Next, we have given a complete review of saturation of an overstable mode
in the two different limits of strong and weak driving force. The literature
for the weak
driving limit, familiar from studies of main sequence or white dwarf
pulsators \citep{1985AcA....35....5D,2001ApJ...546..469W},
is reviewed in detail, and shown not to apply to most physical
situations in which the r-mode instability operates. The strong driving
limit, in which a turbulent cascade forms, has never been applied to
stellar oscillations to our knowledge.
Therefore, the weak turbulence methods in this paper may
find application for amplitude saturation in stars with a driving force
strong enough to parametrically excite many modes, but weak enough that
shocks do not form. In the strong driving limit, we find that the r-mode 
saturates at an energy $E/E_{rotation} \ll \gamma_{gr}/\Omega$.

The consequences of these calculations for neutron star spin evolution are
as follows. First, the time scale for a rapidly rotating ($10^3 \,\rm Hz$)
young neutron star to spin down to a frequency $\nu_{\rm spin}$ by 
gravitational radiation spindown torque is
$t_{\rm spindown} \simeq 2 \times 10^3\, {\rm yr}\, 
(\alpha_e/0.1)^{-1} (\nu_{\rm spin}/ {330\,\rm Hz})^{-11}$. 
Hence, the r-mode can be responsible for an initial, rapid spindown, but
magnetic dipole radiation will dominate for spin frequencies below
$\nu_{\rm spin} \simeq 400\, {\rm Hz}\, (0.1/\alpha_e)^{1/9}\,
B_{12}^{2/9}$.  
%The spindown can also be halted by the windup of
%magnetic field lines.  
Second, in spite of the small saturation amplitude, we find that the gravitational
radiation spin-down torque is still sufficiently large to halt the spin-up by
accretion for neutron stars in LMXB's. Hence, our calculation confirms the validity
of this assumption
by previous investigators. If the viscosity is dominated by normal 
matter, the star will enter into the limit cycle in spin and
temperature discussed by \citet{1999ApJ...517..328L}.  Finally, we
estimate that newly born neutron stars will be visible to $\sim 200
(\alpha_e / 0.1)^{1/2} \, {\rm kpc}$ with enhanced LIGO
interferometers, and LMXB's in the spin down phase of the Levin limit cycle
out to $\sim 90 \, ( \nu_{\rm spin} / 330 \, {\rm
Hz} )^{9/2}  (\alpha_e / 0.1)^{1/2} \, {\rm kpc}$
for $\nu_{\rm spin} \simless 400 \, {\rm Hz}$.

\acknowledgements

It is a pleasure to acknowledge many useful conversations on stellar
oscillations with Yanqin Wu. P.A. would also like to thank
Chris Matzner, Chris Thompson, and Maxim Lyutikov. Part of this work
was completed when two of us (P.A. and S.M) were at the Institute for 
Theoretical Physics at the Spin and Magnetism in Young Neutron Stars
workshop. We thank Lars Bildsten for his gracious
hospitality and for a number of useful conversations, and John
Friedman, Curt Cutler and Peter Goldreich for useful comments on the manuscript.
P.A. is supported by an NSERC fellowship.
This work was supported in part by NSF grants PHY-9900672 and PHY-0084729
at Cornell University. E.E.F. was supported by NSF grants PHY-9722189
and PHY-0140209 and by the Alfred P. Sloan Foundation. S.M. received
support from NSERC.  

% -------------------------------------------------------------

\appendix

\section{cascade solution}
\label{app:cascade}

In this appendix we find scale-free solutions 
to the kinetic equation \ref{eq:kinetic}. The material in this appendix
can be found in texts on weak turbulence theory, e.g.
\citet{1992kstw.book.....Z}, section 3.3. For convenience we use
dimensionless units with $\Eunit=1$ and $\Omega=1$.

We approximate the mode frequency as $\omega = 2 \mu \simeq 2\pi k/n
= s 2\pi |k|/n$, where $s=\pm 1$ is the sign of the frequency. For notational
simplicity, we will use both $k$ and $\mu$ as positive numbers for the
following derivation, using $s$ to take into account the sign.
The resulting expression will then be written in a form
valid for either positive or negative $\mu$.
The sum over modes is then given by
\be
\sum_\alpha & \simeq & \sum_{n=0}^\infty \sum_{k=-n/\pi}^{n/\pi} \sum_{m=-n}^n
\simeq \pi^{-1} \sum_{s=\pm 1}\int_0^\infty dn\, n \int_0^1 d\mu \int_{-n}^n dm.
\ee
Here we used $\mu=\pi k/n$ instead of $k$ since $k$ has an implicit scaling
with $n$.
The coupling coefficients are assumed to be negligible if momentum conservation
is not satisfied, so that 
\be
|\kappa_{\alpha \beta \gamma}|^2 & \simeq & |\bar{\kappa}_{\alpha \beta \gamma}|^2
\delta(s_{\alpha 1}n_\alpha + s_{\beta 1}n_\beta + s_{\gamma 1}n_\gamma)
\delta(s_{\alpha 2}k_\alpha + s_{\beta 2}k_\beta + s_{\gamma 2}k_\gamma)
\delta(m_\alpha + m_\beta + m_\gamma).
\ee
The signs $s_{\alpha 2}$ etc. allow for waves moving in either direction
(see appendix \ref{app:kappa}).
We use the fact that the coupling coefficient
is approximately independent of both $m$ and the sign of $\mu$, and is
separately scale invariant in $n$ and $\mu$ to say
\be
\bar{\kappa}_{\alpha \beta \gamma} & = & \bar{\kappa}(n_\alpha,\mu_\alpha,
n_\beta,\mu_\beta,n_\gamma,\mu_\gamma)
\nonumber \\
\bar{\kappa}(an_\alpha,b\mu_\alpha,an_\beta,b\mu_\beta,an_\gamma,b\mu_\gamma)
& = & a^u b^v \bar{\kappa}(n_\alpha,\mu_\alpha,n_\beta,\mu_\beta,
n_\gamma,\mu_\gamma).
\ee
In appendix \ref{app:kappa} we show that $u=1$ and $v=-2$.
Since the mode frequencies and coupling coefficients are approximately 
independent of
the $m$ quantum numbers, the $m$ dependence can be integrated over giving the
function
\be
A(n_\alpha,n_\beta,n_\gamma) & = & \int_{-n_\alpha}^{n_\alpha} dm_\alpha
\int_{-n_\beta}^{n_\beta} dm_\beta \int_{-n_\gamma}^{n_\gamma} dm_\gamma\,
\delta(m_\alpha + m_\beta + m_\gamma).
\ee
This function is symmetric, and is $A \simeq 4 n_\alpha n_\beta$ in either
the limit $n_\alpha \ll n_\beta \simeq n_\gamma$ or $n_\beta \ll n_\alpha
\simeq n_\gamma$.

Plugging these definitions into eq.\ref{eq:kinetic}, and summing over 
the frequency signs we find
\be
&& \int dm_\alpha\, I_\alpha =  \pi (2/\pi)^2 
\int dn_\beta\, n_\beta dn_\gamma\, n_\gamma d\mu_\beta\, d\mu_\gamma
\ A |\bar{\kappa}_{\alpha \beta \gamma}|^2 \mu_\alpha \mu_\beta \mu_\gamma
\delta(\delta n) \delta(\delta k)
N_\alpha N_\beta N_\gamma
\nonumber \\ 
& \times & 
\left[ \left( \frac{1}{N_\alpha} - \frac{1}{N_\beta} 
- \frac{1}{N_\gamma} \right)\delta(\mu_\alpha - \mu_\beta - \mu_\gamma)
- \left( \frac{1}{N_\beta} - \frac{1}{N_\gamma}
- \frac{1}{N_\alpha} \right)\delta(\mu_\beta - \mu_\gamma -\mu_\alpha)
\right. \nonumber \\ & - & \left.
 \left( \frac{1}{N_\gamma} - \frac{1}{N_\alpha}
- \frac{1}{N_\beta} \right)\delta(\mu_\gamma -\mu_\alpha - \mu_\beta)
 \right].
\label{eq:intidm}
\ee

We attempt to find scale-free inertial range solutions of the form
\be
N & = & N_0 n^{-p}\mu^{-q}.
\ee
After inserting this power law dependence on $n$ and $\mu$ into the 
eq.\ref{eq:intidm},
one can make a change of coordinates in the second and third sets of terms
in the last parenthesis, in order to make them have the same form as the
first term, up to a scaling factor. For instance, in the second set of 
terms let $n_\beta=n_\alpha^2/n_{\beta'}$,
$\mu_\beta=\mu_\alpha^2/\mu_{\beta'}$,
$n_\gamma=n_{\gamma'}n_\alpha/n_{\beta'}$, and
$\mu_\gamma=\mu_{\gamma'}\mu_\alpha/\mu_{\beta'}$.
After simplifying and collecting terms we find
\be
&& \int dm_\alpha\, I_\alpha =  \pi N_0^2 (2/\pi)^2 
\int dn_\beta\, n_\beta dn_\gamma\, n_\gamma d\mu_\beta\, d\mu_\gamma
\ A |\bar{\kappa}_{\alpha \beta \gamma}|^2 \mu_\alpha \mu_\beta \mu_\gamma
\delta(\delta n) \delta(\delta k) \delta(\mu_\alpha - \mu_\beta -\mu_\gamma)
\nonumber \\
& \times &
(n_\alpha n_\beta n_\gamma)^{-p} (\mu_\alpha \mu_\beta \mu_\gamma)^{-q}
\left( n_\alpha^p \mu_\alpha^q - n_\beta^p \mu_\beta^q
- n_\gamma^p \mu_\gamma^q \right)
\nonumber \\
& \times &
\left[ 1 - \left(\frac{n_\alpha}{n_\beta}\right)^{2(p_0-p)}
 \left( \frac{\mu_\alpha}{\mu_\beta} \right)^{2(q_0-q)-1}
- \left( \frac{n_\alpha}{n_\gamma} \right)^{2(p_0-p)}
\left( \frac{\mu_\alpha}{\mu_\gamma} \right)^{2(q_0-q)-1} \right]
\label{eq:cascade1}
\ee
where $p_0=u+3$ and $q_0=v+5/2$.
We can make the scaling of eq.\ref{eq:cascade1} with $n_\alpha$ and
$\mu_\alpha$ explicit by defining $x_\beta=n_\beta/n_\alpha$,
$x_\gamma=n_\gamma/n_\alpha$, $y_\beta=\mu_\beta/\mu_\alpha$,
$y_\gamma=\mu_\gamma/\mu_\alpha$, $\kappa_{\alpha \beta \gamma}=n_\alpha^u
\mu_\alpha^v f_{\alpha \beta \gamma}$, and $\bar{A}=A/n_\alpha^2$.
We find
\be
\int dm_\alpha\, I_\alpha & = & 4 N_0^2 n_\alpha^{2(p_0-p)-2}
\mu_\alpha^{2(q_0-q)-2} \calI(p,q)
\label{eq:cascade2}
\ee
where $\calI(p,q)$ is the dimensionless integral
\be
\calI(p,q) & = & \int dx_\beta\, x_\beta dx_\gamma\, x_\gamma dy_\beta\,
dy_\gamma\,
\bar{A} f^2_{\alpha \beta \gamma} y_\beta y_\gamma
\delta (\delta x) \delta (\delta xy) \delta (\delta y)
(x_\beta x_\gamma)^{-p} (y_\beta y_\gamma)^{-q}
\left( 1 - x_\beta^p y_\beta^q - x_\gamma^p y_\gamma^q \right)
\nonumber \\ & \times &
\left( 1 - x_\beta^{2(p-p_0)} y_\beta^{2(q-q_0)+1}
-  x_\gamma^{2(p-p_0)} y_\gamma^{2(q-q_0)+1} \right).
\label{eq:Ipq}
\ee

Stationary solutions to eq. \ref{eq:cascade1} are now easily found by using
the delta functions to force either the first or second parenthesis to
zero \citep{1992kstw.book.....Z}. 
For instance, using the frequency delta function to set the first
parenthesis to zero gives $N \propto \mu^{-1}$, so that all modes have the
same energy, i.e., thermodynamic equilibrium. Using the momentum delta
functions to set the first parenthesis to zero gives
thermodynamic equilibria with respect to a moving reference frame.
Setting the second parenthesis to zero using the momentum delta functions
gives solutions supporting a constant momentum flux.
We are interested in solutions which
support an energy flux. The frequency delta function can be used to
set the second parenthesis to zero if we let $p=p_0=u+3=4$ and 
$q=q_0=v+5/2=1/2$. We now proceed to show that this solution corresponds
to a flux of energy to small frequency and large wavenumber.

In the inertial range, driving and damping are negligible, and 
the conserved energy flux has components $\calF^n$
and $\calF^\mu$ which satisfy \footnote{ The assumption of local energy
transfer is later shown to be valid by verifying that the flux integrals
converge.}
\be
\omega_\alpha \int dm_\alpha\, I_\alpha + \grad_k \cdot \calFvec_\alpha & = &
\omega_\alpha \int dm_\alpha\, I_\alpha +
\frac{1}{n} \frac{\partial}{\partial n}
\left( n\calF^n_\alpha \right) + \frac{\partial \calF^\mu_\alpha}{\partial \mu}
 =  0.
\label{eq:cascade3}
\ee
Care must be taken in evaluating eq.\ref{eq:cascade2}. For finite
values of $n_\alpha$ and $\mu_\alpha$, taking the limit $(p,q) \rightarrow
(p_0,q_0)$ gives $\int dm_\alpha\, I_\alpha=0$. However, in the vicinity
of $n_\alpha=0$ or $\mu_\alpha=0$ eq.\ref{eq:cascade2} takes on the 
indeterminate form $0/0$, since $\calI \rightarrow 0$ in the numerator and 
either $n_\alpha$ or $\mu_\alpha$ goes to zero in the denominator.
Following Zakharov et.al., we evaluate this expression 
using the delta function representation 
$\mbox{lim}_{\epsilon \rightarrow 0} \epsilon |x|^{\epsilon-1} = 2\delta(x)$
to find
%
% EEF:
% I multiplied this equation across by $\mu_\alpha n_\alpha$ to make
% both sides of the equation be well defined distributions.  
%
\be
\mu_\alpha n_\alpha \int dm_\alpha\, I_\alpha & = & - 4 N_0^2 \left(
\frac{1}{\mu_\alpha} 
\frac{\partial \calI}{\partial p}(p_0,q_0) \delta(n_\alpha)
+ \frac{1}{n_\alpha } \frac{\partial \calI}{\partial q}(p_0,q_0)
\delta(\mu_\alpha)
 \right).
\label{eq:cascade4}
\ee
Hence, the flux can only have a source for $n=0$ or $\mu=0$.
Plugging eq.\ref{eq:cascade4} into eq.\ref{eq:cascade3}, using $\omega = 2\mu$,
and integrating
over $n$ or $\mu$, respectively, gives the two components of the flux
\be
\calF^n & = & 8 N_0^2 n^{-1}\mu^{-1} \frac{\partial \calI}{\partial p}(p_0,q_0)
\nonumber \\
\calF^\mu & = & 8 N_0^2 n^{-2} 
\frac{\partial \calI}{\partial q}(p_0,q_0).
\ee

We now estimate the dimensionless flux integrals, verifying that they 
converge and give have the correct sign. We do this by breaking the 
integration up into two regimes: large daughter mode wavenumber 
($x_\beta \gg 1$) and small daughter mode wavenumber ($x_\beta \ll 1$). 
Although our expansions for the integrand are technically only valid 
in the respective limits, we extend them all the way to $x_\beta \sim 1$.

When the derivatives with respect to $p$ and $q$ are taken in eq.\ref{eq:Ipq},
only the last parenthesis need be differentiated since it gives zero
for $(p,q)=(p_0,q_0)$. The coupling coefficients for the two limits are
given in appendix \ref{app:kappa}. The resulting expressions are
%
% EEF
%
% added footnote noting a restriction on the numerical values of
% $\alpha_n$ and $\alpha_\mu$ requrired for the consistency
% of the cascade picture in Fig. 2.
%
%
%\footnote{Note that the integral curves $n^{\alpha_\mu}
%\mu^{\alpha_n} = $ (const) of the energy flux vector (\ref{eq:calF}) 
%will never reach the shear viscosity dominated regime
%(\ref{eq:shearregime}) in phase space unless $\alpha_n < 2
%\alpha_\mu/3$ (see Fig.\ \ref{fig:cascade}), which is violated by our
%numeric values for $\alpha_n$ and $\alpha_\mu$.  Therefore we suspect
%that a more accurate computation of these constants would yield answers
%satisfying $\alpha_n < 2 \alpha_\mu/3$.} 
\be
\frac{\partial \calI}{\partial p}(p_0,q_0) & \simeq & 
2^{3/2} \int_1^\infty dx_\beta x_\beta^{-2} \ln x_\beta
+ \frac{128}{\pi^3} \int_0^1 dx_\beta x_\beta^{1/2}
\left( \ln x_\beta + 1 \right)
\nonumber \\ & = & 
2^{3/2} + \frac{256}{9\pi^3} \simeq 3.7 \equiv \alpha_n
\ee
and
\be
\frac{\partial \calI}{\partial q}(p_0,q_0) & = & 
- 2^{3/2} \ln{2} \int_1^\infty dx_\beta x_\beta^{-2}
+ \frac{128}{\pi^3} \int_0^1 dx_\beta x_\beta^{1/2} 
\ln x_\beta 
\nonumber \\
& = & -2^{3/2} \ln{2} - \frac{512}{9\pi^3} = -3.8 \equiv - \alpha_\mu.
\ee
We note that the contribution to $\calF^n$($\calF^\mu$)
from both large and small $x_\beta$ are positive (negative).

The energy flux is toward larger $n$ and smaller $\mu$. We now restore the 
sign of $\mu$ in order to have an expression valid for either sign.
The final result for the energy flux is then
\be
\calF^n & = & 8 \alpha_n N_0^2 n^{-1}|\mu|^{-1}
\nonumber \\
\calF^\mu & = & -  8 \alpha_\mu N_0^2 n^{-2} \frac{\mu}{|\mu|}.
\label{eq:Eflux}
\ee
We also note the final answer for the occupation number
\be
N & = & N_0 n^{-4} \mu^{-1/2}
\ee
where $N_0$ is related to the energy flux by eq.\ref{eq:Eflux}.

\section{ analytic estimate of the maximum coupling coefficient}
\label{app:kappa}

Here we give a detailed analytic calculation of the maximum coupling
coefficient. We make the following approximations:
(1) $n \gg m$, the WKB limit; (2) $k_\phi \ll k_R$, which follows from
(1); (3) short wavelength daughter modes with $\mu_\beta \simeq \mu_\gamma
\simeq -\mu_\alpha/2$; (4) an $n=1$ polytropic background
star. For convenience, we use units with $\rstar=M=1$, and we use the
normalization condition $\Eunit=0.5M\rstar^2 \Omega^2$.
In these units, the density profile near the surface is $\rho=\rho_0 \zhat$,
where the central density is $\rho_0=\pi/4$.

We decompose $\psi$ in eq.\ref{eq:efunctions} as a sum of plane waves as
\be
\psi(\xvec) & = & \frac{\psi_0}{(2\pi)^2} 
\left( \frac{\rho}{\rho_0} \sin\theta_1 \sin\theta_2 \right)^{-1/2}
\sum_{s_1,s_2=\pm 1} \exp(i\chi_{s_1 s_2})
\ee
where the WKB phase is
\be
\chi_{s_1 s_2} & =  & p (s_1\theta_1 + s_2 \theta_2) + m\phi + \alpha(s_1+s_2).
\ee
In the small $\mu$ limit, we can 
write $\theta_2 \simeq \pi/2-|\mu| \epsilon$ where $-1 \leq \epsilon 
\leq 1$. Hence the effective wavenumber in the $\theta_2$ direction is
$p |\mu|=\pi |k|$.
From eq.\ref{eq:xi_vs_psi} we find the displacement vector
\be
\xivec(\xvec) & = & \frac{\psi_0}{(2\pi)^2}
\left( \frac{\rho}{\rho_0} \sin\theta_1 \sin\theta_2 \right)^{-1/2}
\sum_{s_1,s_2=\pm 1} \vevec_{s_1 s_2} \exp(i\chi_{s_1 s_2})
\ee
where
\be
\ve_R & = & \frac{i}{1-q^2} \left( k_R + i q k_\phi \right)
\simeq - i\mu^2 k_R
\nonumber \\
\ve_\phi & = & \frac{i}{1-q^2} \left( k_\phi - i q k_R \right)
\simeq - \mu k_R \gg \ve_R
\nonumber \\
\ve_z & = & i k_z \sim \ve_\phi .
\ee
The wavenumber is defined by $\kvec_{s_1 s_2} =
\partial \chi_{s_1 s_2}/ \partial \xvec$
with components found from eq.\ref{eq:ellipoidalcoords} to be
\be
k_R & = & \frac{p(1-\mu^2)^{1/2}}{x_1^2-x_2^2} \left( s_1 x_1 (1-x_2^2)^{1/2}
- s_2 x_2 (1-x_1^2)^{1/2} \right)
\nonumber \\
k_\phi & = & \frac{m}{R} = 
\frac{ m(1-\mu^2)^{1/2} }{ [ (1-x_1^2) (1-x_2^2)]^{1/2} }
\nonumber \\
k_z & = & - s_1 s_2 \frac{ |\mu| }{ (1-\mu^2)^{1/2} } k_R.
\ee

The factor $(x_1^2-x_2^2)/|\mu|$ is the volume element for the bi-spheroidal
coordinates, and becomes zero at coordinate singularities. In the limit
$\zhat \ll 1$ and $\mu \ll 1$, it has the simple form
\be
(x_1^2-x_2^2)^2 & \simeq & (y^2-y_0^2)^2 + \Delta^2
\ee
where $y=\cos\theta$, $y_0 \simeq (\mu^2-2\zhat)^{1/2}$,
and $\Delta=(8\mu^2\zhat)^{1/2}$. This formula shows that there is a 
narrow peak for the wavenumber (for $\zhat \leq \mu^2/2$)
near the singular point $(r,\cos\theta)=(1,\pm |\mu|)$; there is no
pronounced peak for $\zhat > \mu^2/2$ and the integrand decreases
strongly.

Plugging the WKB travelling wave forms into eq.\ref{eq:kappa}
and keeping only the largest terms in the WKB limit gives
\be
\kappa_{\alpha \beta \gamma} & = &  \frac{4\Omega^2 \rho_0}{2\Eunit}
\frac{\psi_{0 \alpha} \psi_{0 \beta}\psi_{0\gamma} }{(2\pi)^6}
\sum_{\vec{s}_\alpha,\vec{s}_\beta,\vec{s}_\gamma}
\int d^3x \frac{\rho}{\rho_0}
\left( \sin\theta_{\alpha 1} \sin\theta_{\alpha 2}
\sin\theta_{\beta 1} \sin\theta_{\beta 2} 
\sin\theta_{\gamma 1} \sin\theta_{\gamma 2} \right)^{-1/2}
\nonumber \\ & \times & 
\left( \frac{\rho_0^3}{\rho_\alpha \rho_\beta \rho_\gamma} \right)^{1/2}
M_{\alpha \beta \gamma} \exp[ i (\chi_\alpha + \chi_\beta + \chi_\gamma )]
\ee
where the matrix element is defined by
\be
M_{\alpha \beta \gamma} & = & \mu_\alpha^2 \kvec_\alpha \cdot \vevec_\beta
\kvec_\alpha \cdot \vevec_\gamma
+ \mu_\beta^2 \kvec_\beta \cdot \vevec_\gamma \kvec_\beta \cdot \vevec_\alpha
+ \mu_\gamma^2 \kvec_\gamma \cdot \vevec_\alpha \kvec_\gamma \cdot \vevec_\beta
\ee
and $\rho_\alpha$ is the cutoff version of the WKB envelope.

To evaluate this expression, we first note that the integral
is rapidly oscillating unless the conservation rules of 
eq. \ref{eq:conservation} are satisfied.
These relations are written
\be
s_{\alpha 1} p_\alpha + s_{\beta 1} p_\beta + s_{\gamma 1} p_\gamma & = & 0
\nonumber \\
s_{\alpha 2} |k_\alpha| + s_{\beta 2} |k_\beta| + s_{\gamma 2} |k_\gamma|
 & = & 0
\nonumber \\
m_\alpha + m_\beta + m_\gamma & = & 0.
\ee
In the limit of large daughter mode wavenumber, we find
$p_\beta \simeq p_\gamma$, $s_{\gamma 1}=-s_{\beta 1}$, $|k_\gamma|
\simeq |k_\beta|$, $s_{\gamma 2}=-s_{\beta 2}$, and $m_\gamma \simeq -m_\beta$;
in other words, the wavevectors of the daughter modes are equal in magnitude
and opposite in direction.
In this limit, the remaining spatially constant phase factor is
\be
\exp[ i (\chi_\alpha + \chi_\beta + \chi_\gamma )]
& = & i^{- \left[ \delta_\alpha(s_{\alpha 1}+s_{\alpha 2})
+ (\delta_\beta-\delta_\gamma)(s_{\beta 1}+s_{\beta 2}) \right] }
\ee
where $\delta=0$ for an even parity mode and $\delta=1$ for an odd parity
mode.
Using the incompressibility condition and momentum
conservation leads to the simplification
$\kvec_\beta \cdot \vevec_\gamma = - \kvec_\alpha \cdot \vevec_\gamma
\simeq  \kvec_\alpha \cdot \vevec_\beta$. Plugging these relations in gives
\be
M_{\alpha \beta \gamma} & = & \frac{1}{2} \mu_\alpha^2 
\kvec_\alpha \cdot \vevec_\beta \left(
- 2 \kvec_\alpha \cdot \vevec_\beta + \kvec_\beta \cdot \vevec_\alpha \right).
\ee
In the limit of small $\mu$ and $k_\phi \ll k_R$ we find
\be
\kvec_\alpha \cdot \vevec_\beta & \simeq &
 -i \frac{\mu_\alpha^2}{4}k_{\alpha R}k_{\beta R}
+ i k_{\alpha z}k_{\beta z}
\nonumber \\
\kvec_\beta \cdot \vevec_\alpha & \simeq & 
-i \mu_\alpha^2k_{\alpha R}k_{\beta R}
+ i k_{\alpha z}k_{\beta z}.
\ee
Performing the spin sums, we find zero for net odd parity, while for
even parity we get
\be
&& \sum_{\vec{s}_\alpha,\vec{s}_\beta,\vec{s}_\gamma}
 M_{\alpha \beta \gamma} 
i^{- \left[ \delta_\alpha(s_{\alpha 1}+s_{\alpha 2})
+ (\delta_\beta-\delta_\gamma)(s_{\beta 1}+s_{\beta 2}) \right] }
 =  
\frac{ \mu_\alpha^6 p_\alpha^2 p_\beta^2 }{ (x_{\alpha 1}^2-x_{\alpha 2}^2)^2
(x_{\beta 1}^2-x_{\beta 2}^2)^2 }
\nonumber \\ & \times & 
\left( 
\left\{ x_{\alpha 1}^2(1-x_{\alpha 2}^2)
+ x_{\alpha 2}^2(1-x_{\alpha 1}^2) \right\}
\left\{ x_{\beta 1}^2(1-x_{\beta 2}^2) 
+ x_{\beta 2}^2(1-x_{\beta 1}^2)\right\}
\right. \nonumber \\ & + & \left.
 4 x_{\alpha 1} x_{\alpha 2} x_{\beta 1} x_{\beta 2}
\left\{ (1-x_{\alpha 1}^2)(1-x_{\alpha 2}^2)(1-x_{\beta 1}^2)
(1-x_{\beta 2}^2) \right\}^{1/2}
\right).
\ee
So far, we have 
\be
\kappa_{\alpha \beta \gamma} & = & \frac{4\Omega^2 \rho_0}{2 \Eunit}
\frac{\psi_{0 \alpha} \psi_{0 \beta}^2}{(2\pi)^6}
 \mu_\alpha^6 p_\alpha^2 p_\beta^2 \calJ
= \frac{p_\alpha \calJ}{4 \pi^{7/2}}
\ee
where the dimensionless integral is
\be
\calJ & = & \int d^3x 
\frac{\rho}{\rho_0} 
\left( \frac{\rho_0^3}{\rho_\alpha \rho_\beta^2} \right)^{1/2}
\left( \sin\theta_{\alpha 1} \sin\theta_{\alpha 2}
\sin^2\theta_{\beta 1} \sin^2\theta_{\beta 2} \right)^{-1/2}
\frac{ 1}{ (x_{\alpha 1}^2-x_{\alpha 2}^2)^2
(x_{\beta 1}^2-x_{\beta 2}^2)^2 }
\nonumber \\ & \times &
\left(
\left\{ x_{\alpha 1}^2(1-x_{\alpha 2}^2)
+ x_{\alpha 2}^2(1-x_{\alpha 1}^2) \right\}
\left\{ x_{\beta 1}^2(1-x_{\beta 2}^2)
+ x_{\beta 2}^2(1-x_{\beta 1}^2)\right\}
\right. \nonumber \\ & + & \left.
 4 x_{\alpha 1} x_{\alpha 2} x_{\beta 1} x_{\beta 2}
\left\{ (1-x_{\alpha 1}^2)(1-x_{\alpha 2}^2)(1-x_{\beta 1}^2)
(1-x_{\beta 2}^2) \right\}^{1/2}
\right).
\ee

To evaluate the dimensionless integral $\calJ$, we first note that if one
attempted to take the $\mu \rightarrow 0$ limit, as for the mode energy,
one would find an integral
\be
\calJ & \propto & \int_0^1 \frac{dR}{R^{1/2}(1-R^2)^2}
\ee
which implies a linear divergence at $R=1$. The divergence implies a strong
dependence of the integral on some characteristic lengthscale near the 
surface.

To proceed with a more detailed calculation, first notice that the 
integrand is sharply peaked near the surface. Below
the daughter mode turning point $\zhat_\beta$, a factor of density
cancels out. The parent mode WKB amplitude is largest for above the
turning point $\zhat \leq \zhat_\alpha$, where the 
cutoff density for the parent mode $\rho_\alpha/\rho_0 \simeq
\zhat_\alpha$. Near the daughter mode singular point 
$\cos\theta=y=|\mu_\beta|$, the bi-spheroidal coordinates become
$x_{\alpha 1} \simeq |\mu_\alpha|$, $x_{\alpha 2} \simeq |\mu_\beta|$,
$x_{\beta 1} \simeq x_{\beta 2} \simeq |\mu_\beta|$. A similar expression
can be found near the parent mode singular point.
The integral $\calJ$ then becomes
\be
\calJ & = &
16\pi \zhat^{-1/2}_\alpha
\int_{\zhat_\beta}^{\zhat_\alpha} d\zhat
\int_{-1}^{-1} \frac{dy}{(y^2-y_{\alpha 0}^2)^2 + \Delta_\alpha^2}
+  4\pi \zhat^{-1/2}_\alpha
\int_{\zhat_\beta}^{\zhat_\alpha} d\zhat
\int_{-1}^{-1} \frac{dy}{(y^2-y_{\beta 0}^2)^2 + \Delta_\beta^2}
\nonumber \\ & = & 
16\pi \zhat^{-1/2}_\alpha
\int_{\zhat_\beta}^{\zhat_\alpha} d\zhat \frac{\pi}{\Delta_\alpha y_{\alpha 0} }
+ 4\pi \zhat^{-1/2}_\alpha
\int_{\zhat_\beta}^{\zhat_\alpha} d\zhat \frac{\pi}{\Delta_\beta y_{\beta 0} }
\nonumber \\ & = & 
\frac{4\sqrt{2}\pi^2}{\zhat^{1/2}_\alpha \mu_\alpha^2}
\int_{\zhat_\beta}^{\zhat_\alpha} \frac{\zhat}{\zhat^{1/2} }
+ \frac{4\sqrt{2}\pi^2}{\zhat^{1/2}_\alpha \mu_\alpha^2}
\int_{\zhat_\beta}^{\zhat_\alpha} \frac{\zhat}{\zhat^{1/2} }
\simeq \frac{16 \sqrt{2} \pi^2}{\mu_\alpha^2 }.
\ee
The final result for the coupling coefficient for nearly identical,
short lengthscale daughter modes is then
\be
\kappa_{\alpha \beta \gamma} & = & 
= \frac{p_\alpha }{4 \pi^{7/2}}  \frac{16 \sqrt{2} \pi^2}{\mu_\alpha^2 }
= 1.02 n_1 \mu_1^{-2}.
\label{eq:kappalarge}
\ee

In appendix \ref{app:cascade} when finding the cascade solution, one also
needs the coupling coefficient in the limit that one of the daughter modes
has a small wavenumber. 
In section \ref{sec:epconservation} we found that the approximate solutions
in the limit $n_\gamma \ll n_\alpha \sim n_\gamma$ are: 
$n_\gamma \simeq n_\alpha + n_\beta$, $k_\gamma \simeq k_\alpha$,
$k_\beta \simeq k_\alpha(n_\beta/n_\alpha)^2$, $\mu_\gamma \simeq -\mu_\alpha$,
and $\mu_\beta \simeq \mu_\alpha(n_\beta/n_\alpha) \ll \mu_\alpha$. 
Since the method is exactly
the same as the detailed calculation already given, we merely quote
the answer for the maximum coupling coefficient to be
\be
\kappa_{\alpha \beta \gamma} & = & \left(\frac{8}{\pi^3}\right)^{1/2}
\frac{p_\alpha}{|\mu_\alpha \mu_\beta|} \left( \delta_O + 
\delta_S \mu_\beta/2\mu_\alpha \right).
\label{eq:kappasmall}
\ee
Here $\delta_O=1$ if modes $\alpha $ and $\gamma$ have the opposite parity,
and $\delta_O=0$ otherwise. Similarly, $\delta_S=1$ if modes 
$\alpha $ and $\gamma$ have the same parity,
and $\delta_S=0$ otherwise.

%\bibliographystyle{apj}
%\bibliography{apj-jour,ref}

\end{document}